\documentclass[pre,amsmath,amssymb,aps,twocolumn,nofootinbib]{revtex4}
\usepackage{graphicx}
\usepackage{feynmp,color}
\usepackage{times}
\usepackage{float}
\usepackage[caption = false]{subfig}

\usepackage{xcolor}

\newcommand{\highlight}[1]{#1}

\newcommand{\Mathematica}[1]{}

\newcommand{\Eq}[1]{Eq.~(\ref{#1})}
\newcommand{\Eqs}[1]{Eqs.~(\ref{#1})}
\newcommand{\eq}[1]{(\ref{#1})}

\newcommand{\half}{\frac12}

\newcommand{\bea}{\begin{eqnarray}}
\newcommand{\eea}{\end{eqnarray}}
\newcommand{\beq}{\begin{equation}}
\newcommand{\eeq}{\end{equation}}
\newcommand{\be}{\begin{equation}}
\newcommand{\ee}{\end{equation}}

\newcommand{\rme}{\mathrm{e}}
\newcommand{\rmd}{\mathrm{d}}
\newcommand{\nn}{\nonumber}
\renewcommand{\epsilon}{\varepsilon}
\newcommand{\nott}[1]{}
\newcommand{\Fig}[1]{\includegraphics[width=\columnwidth]{./#1}} \newcommand{\fig}[2]{\includegraphics[width=#1\columnwidth]{./#2}}
\newcommand{\cfig}[2]{\includegraphics[width=#1]{./#2}}

\newlength{\bilderlength}

\graphicspath{{figures/},{}}

\renewcommand{\paragraph}{\subsubsection*}

\begin{document}

\title{The Spatial Shape of Avalanches}
\author{Zhaoxuan Zhu and  Kay J\"org Wiese}
  \affiliation{CNRS-Laboratoire de Physique Th\'eorique de l'Ecole Normale Sup\'erieure, PSL Research University, Sorbonne Universit\'es, UPMC, 24 rue Lhomond, 75005 Paris, France. 
}

\begin{abstract}
In  disordered elastic systems,   driven by displacing a parabolic confining potential adiabatically slowly, all advance of the system is in bursts, termed avalanches. Avalanches have a finite extension in time, which is much smaller than the waiting-time between them. 
 Avalanches also have a finite extension $\ell$ in space, i.e.\ only a part of the interface of size $\ell$ moves during an avalanche. Here we study their spatial shape $\left< S(x)\right>_{\ell}$ given     $\ell$, as well as  its fluctuations encoded in the second  cumulant $\left< S^{2}(x)\right>_{\ell}^{{\rm c}}$. We establish  scaling relations governing the behavior close to the boundary.  
 We then give analytic results for the Brownian force model, in which the microscopic disorder for each degree of freedom is a random walk. Finally, we confirm these results with numerical simulations. To do this properly
 we elucidate the influence of discretization  effects, which also confirms the assumptions entering into the scaling ansatz. This allows us to  reach the scaling limit already  for avalanches of moderate size.
 We find excellent agreement for the  universal shape, its fluctuations, including all  amplitudes. 
 \end{abstract}

\maketitle

\section{Introduction}
Many physical systems in the presence of disorder, when driven adiabatically slowly,  advance in abrupt bursts, called avalanches. The latter can be found in the domain-wall motion in soft magnets  \cite{DurinZapperi2000}, in fluid contact lines on a rough surface \cite{LeDoussalWieseMoulinetRolley2009}, slip instabilities leading to earthquakes on geological faults, or in fracture experiments \cite{BonamyPonsonPradesBouchaudGuillot2006}. In magnetic systems they are   known as Barkhause noise \cite{Barkhausen1919,DurinZapperi2006b,DurinBohnCorreaSommerDoussalWiese2016}. 
In some experiments \cite{LeDoussalWieseMoulinetRolley2009}, but better in numerical simulations \cite{RossoLeDoussalWiese2009a,AragonKoltonDoussalWieseJagla2016,FerreroFoiniGiamarchiKoltonRosso2017} it can be seen that avalanches have a well-defined extension, both in space, as in time. In theoretical models, this is achieved without the introduction of a short-scale cutoff. 
This is non-trivial: The velocity in an avalanche, i.e.\ its {\em temporal shape},  could well decay exponentially in time, as is the case in magnetic systems in presence of eddy currents \cite{DobrinevskiLeDoussalWiese2013,ZapperiCastellanoColaioriDurin2005}. However it can be shown that    generically an avalanche stops abruptly. In a field-theoretic expansion   \cite{DobrinevskiLeDoussalWiese2014a} the velocity of the center of mass inside an avalanche of duration $T$ was shown to be   well approximated by 
\be \textstyle \label{1}
\left< \dot u(t=x T)\right>_T\simeq [ Tx(1-x)]^{\gamma-1} \exp\!\left( {\cal A}\left[\frac12-x\right]\right)
\ ,
\ee
where $0<x<1$. 
 The exponent $\gamma=(d+\zeta)/z$ is given by the two independent exponents at depinning, the roughness $\zeta$ and the dynamical exponent $z$.  The asymmetry ${\cal A}$ is  negative for $d$ close to $d_{\rm c}$, i.e.\ ${\cal A}\approx - 0.336 (1-d/d_{\rm c})$ skewing the avalanche towards its end, as observed in numerical simulations in $d=2$ and $3$ \cite{LaursonPrivate}. In one dimension, the asymmetry  is positive \cite{LaursonIllaSantucciTallakstadyAlava2013}. While more precise theoretical expressions are available \cite{DobrinevskiLeDoussalWiese2014a}, an experimental or numerical verification of these finer details is difficult, and currently lacking.  

In this article, we analyze not the temporal, but the spatial shape $\left< S(x)\right>_\ell$ of an avalanche of extension $\ell$. To define this shape properly, it is, as  for the temporal shape,   important that an avalanche has   well-defined endpoints in space, and a well-defined extension $\ell$.

Let us start to review where the theory on avalanches stands. 
The  systems mentioned above can efficiently be modeled by an elastic interface driven through a disordered medium, see \cite{DSFisher1998,WieseLeDoussal2006} for a review of basic properties.
The energy functional for such a system has the   form 
\be\label{energy}
{\cal H}[u] = \int_x \frac12 [\nabla u(x)]^2 + \frac {m^2}{2}[u(x)-w]^2 +V\big(x,u(x)\big) \ .
\ee
The term $V(x,u)$ is the disorder potential, correlated as $\overline{V (x,u) V(x',u')} = \delta^d (x-x')R(u-u') $. 
The term proportional to $m^2$ represents a confining potential centered at $w$. Changing $w$ allows to study avalanches, either in the {\em statics} by finding the minimum-energy configuration; or in the dynamics,  at {\em depinning}, by studying the associated Langevin equation (usually at zero temperature)
\bea\label{Langevin}
\gamma \partial_t u(x,t) &=& - \frac{\delta {\cal H}[u]}{\delta u(x)}
\bigg|_{{u(x)=u(x,t)}} \\
&=& \nabla^2 u(x,t) - m^2 [u(x,t)-w] + F\big(x,u(x,t)\big). \nn
\eea
The random force $F(x,u)$ in Eq.~(\ref{Langevin}) is related to the random potential $V(x,u)$ by 
$
F(x,u) = - \partial_u V(x,u)
$. It has correlations $\overline{F(x,u),F(x',u')} = \delta^d (x-x') \Delta(u-u')$, related to the correlations of the disorder-potential via $\Delta(u) = -R''(u)
$. 
To simplify notations, we rescale time by $t\to t/\gamma$, which sets the coefficient $\gamma=1$ in Eq.~(\ref{Langevin}).

It is important to note that $\partial_t u(x,t)\ge 0$, thus the movement is always forward (Middleton's theorem \cite{Middleton1992}). This property is important for the avalanche dynamics, and for a proper construction of the field-theory. 
 Much progress was achieved in this direction over the past years, thanks to a powerful method, the Functional Renormalization group (FRG). It was first applied to a precise estimation of the critical exponents \cite{DSFisher1986,NattermannStepanowTangLeschhorn1992,NarayanDSFisher1992b,ChauveLeDoussalWiese2000a,LeDoussalWieseChauve2002,LeDoussalWieseChauve2003}.  
Later it was realized and verified in numerical simulations that the central object of the field theory is directly related to  the correlator of the  center-of-mass fluctuations, both in the statics \cite{MiddletonLeDoussalWiese2006} and at depinning \cite{RossoLeDoussalWiese2006a}. 

To build the field-theory of avalanches, one first  identifies the upper critical dimension, $d_{c}=4$ for standard (short-ranged) elasticity as in \Eq{energy}, or $d_{c}=2$ for long-ranged elasticity. For depinning, it was proven that at this upper critical dimension, the relevant  (i.e.\ mean-field) model is the Brownian Force model (BFM):  an elastic manifold 
with Langevin equation (\ref{Langevin}),  in which the random force experienced by each degree of freedom has the statistics of a random walk, i.e.\ \cite{LeDoussalWiese2012a,LeDoussalWiese2010b,DobrinevskiLeDoussalWiese2011b}
\be\label{DeltaFBM}
\Delta(0)- \Delta(u-u') = \sigma |u-u'|\ .
\ee
The BFM   then serves as the starting point of a controlled $\epsilon$-expansion, $\epsilon = d_{\rm c}-d$, around the upper critical dimension. This is relevant  both for equilibrium, i.e.\ the statics \cite{LeDoussalWiese2011b,LeDoussalWiese2009a,RossoLeDoussalWiese2009a,LeDoussalWiese2008c,LeDoussalMiddletonWiese2008} as at depinning  \cite{DobrinevskiLeDoussalWiese2013,DobrinevskiPhD}. Results are now available for the avalanche-size distribution, the distribution of durations, and the temporal shape, both at fixed duration $T$ as given in Eq.~(\ref1), and at fixed size $S$.

 Much less is known   about the spatial shape, i.e.\ the expectation of the total advance inside an avalanche as a function of space, given a total extension $\ell$. To simplify our considerations and notations, consider  dimension $d=1$. 
 There  this is a function $\left<S(x) \right>_{\ell}$, vanishing for $|x|> \frac \ell2$. 
 
 Most results currently available were obtained for the BFM. A first important step was achieved in   Ref.\ \cite{ThieryLeDoussalWiese2015}. Starting from an exact functional for the probability to find an avalanche of shape $S(x)$ (reviewed in section \ref{s:large-avalanches}), a saddle-point analysis permitted to obtain the shape for avalanches of size $S$,  with a   large aspect ratio $S/\ell^{4}\gg 1$. It was shown that in this case the mean avalanche shape grows as $\left< S(x)\right>_{\ell,S}\sim (x-\ell/2)^{4}$ close to the (left) boundary. A subsequent expansion in $\frac{\ell^{4}}S $ allowed the authors to include corrections for smaller sizes. This did not change the  scaling close to the boundary. 
 
 We   believe that this scaling does not pertain to generic avalanches\footnote{This is contrary to the claim made in Ref.~\cite{ThieryLeDoussalWiese2015}, that in the BFM also for generic avalanches the scaling exponent close to the boundary is $4$. We   show  in appendix \ref{a:replot} by reanalyzing the data  of \cite{ThieryLeDoussalWiese2015} that they   favor an exponent 3 instead of 4, in agreement with our results (\ref{gofx}) and (\ref{zetaBFM}).}: Avalanches which have an extension $\ell \ll L_{m}=m^{-1}$, i.e.\ the infrared cutoff set by the confining potential  in Eqs.~(\ref{energy}) or (\ref{Langevin}), should obey the scaling form
\be\label{ansatz}
\left< S(x)\right>_\ell =  \ell^\zeta g(x/\ell)
\ ,
\ee
where $g(x)$ is non-vanishing in the interval $[-1/2,1/2]$. Integrating this relation over space yields 
$
S\sim \ell^{d+\zeta}, 
$
the canonical scaling relation between size and extension of avalanches, confirming the ansatz (\ref{ansatz}). 

We now want to deduce how $g(x)$ behaves close to the boundary. For simplicity of notations, we write our argument for the left boundary in $d=1$. Imagine the avalanche dynamics for a discretized representation of the system. The avalanche starts at some point, which in turn triggers avalanches of its neighbors, a.s.o. This will lead to a shock front   propagating outwards from the seed to the left and to the right. As long as the elasticity is local as in \Eq{energy},  the dynamics of these two shock-fronts is local: If one conditions on the position of the $i$-th point away from the boundary, with $i$ being much smaller than the total extension $\ell$ of the avalanche (in fact, we only need that the avalanche started right of this point), then we expect that the joint probability distribution for the advance of points $1$ to $i-1$  depends on $i$, but is independent of the size $\ell$. 
Thus we expect that   {\em in this discretized model} the shape $\left< S(x-r_1)\right> $ close to the left boundary $r_1$ is  independent of $\ell$. Let us call this the {\em boundary-shape conjecture}. We will verify later in numerical simulations that it indeed holds.  

\begin{figure}[t]
\Fig{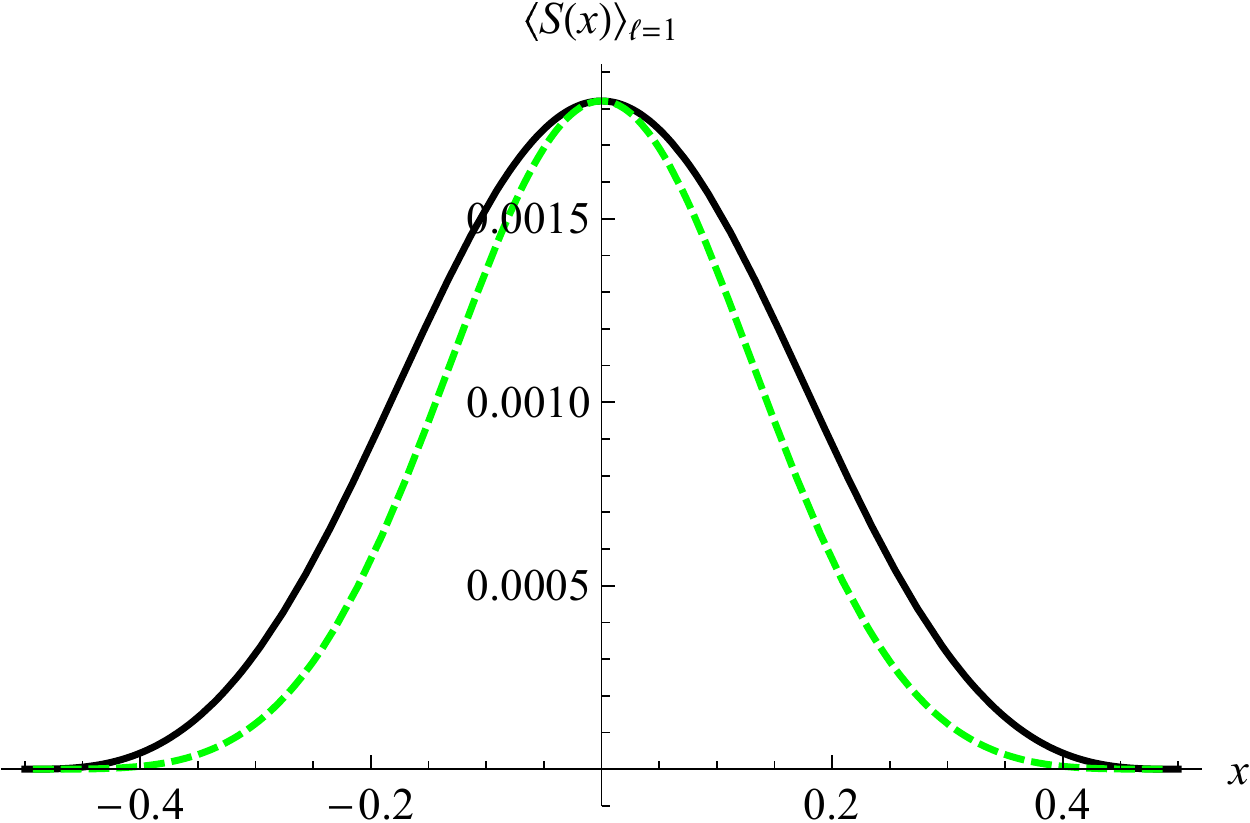}
\caption{The avalanche shape $\left<S(x) \right> _{\ell=1}$ for $\sigma=1$.  The green dotted line is the shape obtained for  avalanches with a large aspect ratio $S/\ell^{4}$ at fixed $S$ and $\ell$ in \cite{ThieryLeDoussalWiese2015}, rescaled to the same height at $x=0$.}
\label{f:spatial-shape}
\end{figure}

Let us now turn to avalanches of large size $\ell$, so that we are in the continuum limit studied in the field theory. 
Our conjecture then implies that the shape $\left< S(x-r_1)\right>$ measured from the left boundary $r_1=-\ell/2$, is independent of $\ell$. In order to cancel the $\ell$-dependence in Eq.~(\ref{ansatz}) this in turn implies that 
\be \label{gofx}
g(x-1/2) = {\cal B} \times  (x-1/2)^\zeta\ , 
\ee
with some  amplitude ${\cal B}$. For the Brownian force model in $d=1$, the roughness exponent is 
\be\label{zetaBFM}
\zeta_{\rm BFM} = 4-d = 3\ .
\ee
We will show below that in the BFM the   amplitude $\cal B$ is given by \be\label{B}
{\cal B} = \frac\sigma{21}\ .
\ee
We further show that the function $g(x)=\left< S(x) \right>_{\ell=1}$ for the BFM can be expressed in terms of a Weierstrass-${\cal P}$ function and its primitive, the Weierstrass-$\zeta$ function, see Eqs.~(\ref{expS(x)}),  (\ref{50}), and (\ref{Fdef}). This function is plotted on figure \ref{f:spatial-shape} (solid, black).
For comparison, we also give the shape for avalanches with a large aspect ratio $S/\ell^4$ \cite{ThieryLeDoussalWiese2015}, rescaled to the same peak amplitude (green dashed). The two shapes are significantly different.  

We  would   like to mention the study \cite{ThieryLeDoussal2016} of avalanche shapes, conditioned to start at a given seed, and having total size $S$. This particular conditioning renders the solution in the BFM essentially trivial: the spatial dependence becomes that of diffusion, so the final result is the center-of-mass velocity folded with the diffusion propagator. The advantage of this approach is that one can relatively simply include perturbative corrections in $4-d$ away from the upper critical dimension. A   shortcoming is that the such defined averaged shape is far from sample avalanches seen in a simulation: Especially, one of the key features, namely the finite extension of each avalanche encountered in a simulation, is lost. When applied to experiments, it is furthermore questionable whether one will be able to identify the seed of an avalanche. 
For these reasons, we will develop below the theory of avalanches with given spatial extension $\ell$.

\section{The probability of a given spatial avalanche shape}
\label{s:large-avalanches}
Here we review some basic results of Ref.~\cite{ThieryLeDoussalWiese2015} for the Brownian Force Model. Suppose that the interface is at rest in configuration $u_1(x) = u(x,t_1)$, and then an avalanche occurs which brings it to   configuration $u_2(x)=u(x,t_2)$. Denote $S(x):=u_2(x)-u_1(x)$ the total advance at point $x$, which we call the {\em spatial shape} of the avalanche.

We start with a simplified derivation of the key formula of Ref.\ \cite{ThieryLeDoussalWiese2015}, given below in Eq.~(\ref{PofS}). To this aim, we write  the MSR action for the dynamics of the interface, obtained from a time derivative of \Eq{Langevin},  as \cite{LeDoussalWiese2011a,LeDoussalWiese2012a,DobrinevskiLeDoussalWiese2011b}
\begin{eqnarray}\label{9}
&&\!\!\!\rme^{-{\cal S}[\tilde u, \dot u]} =\\
&& \!\!\!\rme^{\int_{x,t} \tilde u(x,t) \left[ - \partial_t\dot u(x,t) + \nabla^2 \dot u(x,t) - m^2 \dot u(x,t) +\partial_t F(x,u(x,t)) + \partial_t f(x,t)\right] }. \nn
\end{eqnarray}
There are no avalanches without driving, and the last term has been added for this purpose.  We want to drive the system with a force kick at $t=0$, i.e.\ \be
\label{driving}
f(x,t) = \delta (t) w(x)\ .
\ee
Note that compared to the notations in Refs.~\cite{LeDoussalWiese2011a,LeDoussalWiese2012a,DobrinevskiLeDoussalWiese2011b} we have absorbed a factor of $m^2$ into $w$: Here as in Ref.~\cite{DelormeLeDoussalWiese2016} it is a kick in the force, there it is a kick in the displacement. Our choice is made so that the  limit of $m\to 0$ can be taken later. 

To obtain static quantities (as the avalanche-size distribution), one can use a time-independent response field  $\tilde u(x,t)=\tilde u(x)$  \cite{LeDoussalWiese2012a,DobrinevskiLeDoussalWiese2011b}. Integrating over times from $t_1$ before the avalanche to $t_2$ after the avalanche, and using that the interface is at rest at these two moments, yields\footnote{This does not take into account the change of measure from $\prod_{t}\rmd \dot u(x,t)$ to $\rmd S(x)$, and similarly for $\tilde u(x,t)$. Our simplified derivation thus misses an  additional global factor in Eq.~(50) of \cite{ThieryLeDoussalWiese2015}. Especially, the result (\ref{PofS}) is incorrect for a single degree of freedom. On the other hand, integrating Eq.~(\ref{11}) over $S(x)$ still gives the correct instanton equation (\ref{basic}), which can be derived independently from this argument, see e.g.\ \cite{DelormeLeDoussalWiese2016}.}
\begin{eqnarray}
&&\rme^{-{\cal S}[\tilde u, u]}= \\
&&\qquad  \rme^{\int_{x} \tilde u(x) \left[w(x)+ \nabla^2 S(x) - m^2 S(x) + F(x,u_2(x)) - F(x,u_1(x))\right]}
\ .\nn
\end{eqnarray}
Averaging over disorder, using $\overline{ F(x,u) F(x',u')} = \delta^d(x-x')\Delta(u-u')$, we obtain 
\begin{eqnarray}\label{11}
&&\!\!\!\overline{ \rme^{-{\cal S}[\tilde u, S]} }\\
&&\qquad = \rme^{\int_{x} \tilde u(x) \left[w(x)+\nabla^2 S(x) - m^2 S(x)\right] +\tilde u(x)^2 [ \Delta(0)-\Delta(S(x)) ]}  .\nn
\end{eqnarray}
Integrating over $\tilde u(x)$ yields 
\begin{eqnarray}
&& \int {\cal D}[\tilde u] \,\overline{ \rme^{-{\cal S}[\tilde u, u]} } \simeq  \prod_{x}\frac1{\sqrt{\Delta(0)-\Delta(S(x))}} \times \nn\\
&& \qquad\times \exp  \!\left({-\frac{\left[ \int_{x} w(x)+ \nabla^2 S(x) - m^2 S(x)\right]^2}{4 [ \Delta(0)-\Delta(S(x))]}}\right).    ~~~~~~~~
\end{eqnarray}
This formulas is a priori exact for any disorder correlator $\Delta(u)$. 
For the BFM $\Delta(0)-\Delta(u) \equiv \sigma |u|$. Thus we obtain  upon  simplification in the limit of $w(x) \to 0$ \cite{ThieryLeDoussalWiese2015}
\begin{equation}\label{PofS}
\mbox{proba}\big[S(x)\big] \simeq \prod_{x}\frac1{\sqrt{S(x)}}\, \rme^{-\int_x  \frac{ m^4 S(x)}{4 \sigma} + \frac{[\nabla^2 S(x)]^2}{4\sigma S(x)}}\ .
\end{equation}
Changing variables to $\phi(x):= \sqrt{S(x)}$ eliminates the factor of $\prod_{x}S(x)^{-1/2} $. 
A saddle point for avalanches with a   large aspect ratio $S/\ell^4$, where $S$ is the  avalanches size and $\ell$ its spatial extension,  can be obtained by varying w.r.t.\ $\phi(x)$. The  solution of this saddle-point equation is plotted on figure \ref{f:spatial-shape} (green dashed line), where it is confronted to the shape for generic avalanches (black) to be derived  later. See also figure \ref{f:data-Thimotee} for a numerical validation of the saddle-point solution in reference \cite{ThieryLeDoussalWiese2015}.

\section{The expectation of $S(x)$ in an avalanche extending from $-\ell/2$ to $\ell/2$}

\subsection{Generalities}

We consider avalanches in the BFM  in $d=1$ dimensions. 
To this aim, we start from \Eq{11}, using the correlator \eq{DeltaFBM}. This yields 
\begin{equation}\label{14}
\rme^{-{\cal S}_{\rm FBM}[\tilde u, S]} = 
\rme^{\int_{x} \tilde u(x) \left[w(x)+\nabla^2 S(x) - m^2 S(x)\right]+ \sigma  \tilde u(x)^2 S(x)}.
\end{equation}
We now wish to evaluate the generating function for avalanche sizes 
\bea\label{basic}
 \tilde {\cal P}\big[ \lambda(x)\big] &: =&
\overline {\rme^{\int_x\lambda(x) S(x)}} \nn\\
&=&  \int {\cal D}[S]\, {\cal D}[\tilde u]\, \rme^{ \int_x\lambda(x) S(x) -{\cal S}_{\rm FBM}[\tilde u, S]}
\ .
\eea
The crucial remark is that $S(x)$ appears  {\em linearly} in the exponential; thus integrating over $S(x)$ enforces that
 $\tilde u(x)$ obeys the differential equation \cite{DelormeLeDoussalWiese2016}
\be \label{6}
\tilde u''(x) - m^2\tilde u(x) +\sigma \tilde u(x)^2 =  -\lambda(x)\ .
\ee
This is an instanton equation. Suppose we have found its solution, which for simplicity  we also denote   $\tilde u(x)$.  Then \Eq{basic} simplifies considerably to \cite{DelormeLeDoussalWiese2016}
\be\label{basic2}
 \tilde {\cal P}\big[ \lambda(x)\big] : =
\overline {\rme^{\int_x\lambda(x) S(x)}} 
=\rme^{\int_x   w(x) \tilde u(x) }\ .
\ee
In Ref.~\cite{DelormeLeDoussalWiese2016}, a solution for $\lambda(x)$ in the form 
\be\label{lambda(x)}
\lambda(x) = -\lambda_1 \delta(x-r_1) -\lambda_2 \delta(x-r_2)
\ee
was given in the limit of $\lambda_{1,2}\to \infty $.  This solution ensures that if the interface has  moved at positions $r_1$ or $r_2$, the expression  $ {\rme^{\int_x\lambda(x) S(x)}}$ is 0; otherwise it is 1. The probability that the interface has not moved at these two positions $r_1$ and $r_2$ thus is 
\be \label{probr1r2}
\tilde {\cal P}_{r_1,r_2}=\rme^{\int_x w(x) \tilde u(x)}\ .
\ee
We now consider driving at   $x$  between the two points $r_1$ and $r_2$. In order that the probability (\ref{probr1r2}) decreases for an increase in the driving at   $x$, we need that 
\be\label{constraint}
\tilde u(x)< 0 \ , \qquad r_1<x<r_2\ .
\ee
This helps us to select the correct solution, see appendix \ref{a:instanton-solution-recall}. 
Call $\tilde u_0(x)$ this solution. According to \cite{DelormeLeDoussalWiese2016}, it reads
\be\label{8}
\highlight{\tilde u_0(x) = \frac{1}{(r_2-r_1)^2}\, f\!\left(\frac{2x-r_1-r_2}{2(r_2-r_1)} \right) }
\ .
\ee
Its extension is 
\be
\ell = r_2-r_1\ .
\ee
It further depends on the dimensionless combination  $\frac{\ell}{L_m} = \ell m$.
In the massless limit, i.e.\ for
\be
\frac{\ell}{L_m} = \ell m \ll 1\ ,
\ee
the function $f(x)$ satisfies Eq.~(\ref{6}) for $m=0$, i.e. 
\be\label{44}
f''(x) + f(x)^2 = 0\ .
\ee
This solution diverges with a quadratic divergence at $x=\pm 1/2$. We   review in appendix \ref{a:instanton-solution-recall} its construction. We see there that it is a {\em negative-energy} solution with energy $-\bar {\cal E}_1$, where 
\be
\bar{\cal E}_1=          \frac{8 \pi ^3 \Gamma
   \left(\frac{1}{3}\right)^6}{3 \Gamma
   \left(\frac{5}{6}\right)^6}\ .
\ee
It reads
\be \ 
f(x) = - 6 \,{\cal P}\!\left(x+1/2; g_2=0,g_3 = \frac{\Gamma \left(\frac{1}{3}\right)^{18}}{
   (2\pi) ^6}\right) \label{50}
\ .
\ee
The function ${\cal P}$ is the Weierstrass $\cal P$ function. The parameter $g_3$ satisfies 
\be
\highlight{ g_3 \equiv  \frac{\bar {\cal E}_1}{18}}\ ,
\ee
and the solution respects the constraint (\ref{constraint}).
For later simplifications we note the following relations
\bea 
\label{45}
&& \highlight{\frac23 f^3(x)+f'(x)^2 =-36  g_3 \equiv  -2 {\bar {\cal E}_1} \ .} \\
\label{46}
&& \highlight{\frac 23 f(x)f''(x)-f'(x)^2 = 36 g_3 = 2 \bar {\cal E}_1\ .}
\eea

\subsection{Driving}
\label{s:driving}
Let us now specify the driving function $w(x)$ introduced in Eq.~(\ref{driving}). There are two main choices:
\begin{itemize}
\item[(i)] uniformly distributed random seeds (random localized driving)
\be
w(x) = w\, \delta (x-x_{\rm s})\ .
\ee
Here we first calculate the observable at hand, and finally  average, i.e.\ integrate, over the seed position $x_{\rm s}$. In a numerical experiment, one can take a random permutation of the $N$ degrees of freedom, and then apply a kick to each of them in the chosen order.  
\item[(ii)] uniform driving 
\be
w(x) = w\ .
\ee
\end{itemize}
As we wish to work at first non-vanishing order in $w$, this makes {\em almost} no difference. Indeed, in Eq.~(\ref{basic2}) we formally have for both driving protocols to leading order in $w$
\be\label{33}
\overline {\rme^{\int_x\lambda(x) S(x)}-1} = \rme^{\int_x   w (x) \tilde u(x) }-1 \to w  \int_x \tilde u(x) \ .
\ee
There is however one caveat: If $\tilde u(x)\to -\infty$,  as is the case for solution (\ref{8}) at $x=r_{1,2}$, then, for localized driving, points $x$ around these singularities are suppressed, and the corresponding points have to be taken out of the integral. On the other hand, for uniform driving, the middle integral in \Eq{33} simply vanishes. In that case one has to regularize the solution, i.e.\ work at finite $\lambda_{1,2}$, then take $w\to0$, and only at the end take the limit $\lambda_{1,2}\to \infty$.  
 According to appendix \ref{a:instanton-solution-recall}, working at finite $\lambda_{1,2}$ is equivalent to cutting out a piece of size  $x_0$ around the singularity, with $x_0$   given by Eq.~(\ref{A3}). Thus, effectively,  driving is restricted to the interval $[r_1+x_0,r_2-x_0]$ slightly smaller than the the full interval $[r_1,r_2]$.

For conceptual clarity, and simplicity of presentation,  we will work with uniformly distributed random seeds (random localized driving) below. 
The idea to keep in mind is that in the limit $w\to0$, the driving only triggers the
avalanche, but after the avalanche starts, its subsequent dynamics is independent of the driving.  As a result, the avalanche shape is   independent
of the driving   and we can choose the most  convenient driving.

\subsection{Strategy of the calculation}
We now want to construct perturbatively a solution of \Eq{6} at $m=0$,  and $\sigma=1$. 
i.e. 
\be\label{34}  
\tilde u''(x)  +\tilde u(x)^2 =  -\lambda(x)\ , 
\ee 
with 
\begin{align} \label{30}
& \lambda(x) = -\lambda_1 \delta(x-r_1) -\lambda_2 \delta(x-r_2) + \eta \delta(x-x_{\rm c})\ ,\\
& \lambda_1,\lambda_2 \to \infty\ .
\end{align}
We are interested 
in the limit of vanishing $\eta$, i.e.\ at first and second order in $\eta$. This instanton solution  will have the   form  
\be\label{31}  
\tilde u(x) = \tilde u_0(x) + \eta \tilde u_1(x)+ \eta^2 \tilde u_2(x)+ ...
\ .
\ee
It will be continuous, but non-analytic at $x=x_{\rm c}$, see Fig~\ref{f-lambda-plot-order-1}. 

Let us  reconsider \Eq{basic2}, i.e.\
$
\overline {\rme^{\int_x\lambda(x) S(x)}} 
=\rme^{\int_x   w(x) \tilde u(x) }
$.
Its l.h.s.\ can be written as 
\bea \label{38}
\overline {\rme^{\int_x\lambda(x) S(x)}} &=&\int_{r_1}^{x_{\rm c}}\rmd r_{\rm left} \int^{r_2}_{x_{\rm c}}\rmd r_{\rm right} \int_0^\infty {\rmd} S(x_c)\,\nn\\
&& \qquad   \rme^{\eta S(x_c)} {\cal P}(S(x_c),r_{\rm left},r_{\rm right})\ ,\qquad \qquad 
\eea
where ${\cal P}(S(x_c),r_{\rm left},r_{\rm right}) $ is the joint probability that the avalanche  has advanced by $S(x_{\rm c})$ at $x_c$, and  that it extends from  $r_{\rm left}$ to  $r_{\rm right}$,  with $r_1<r_{\rm left}<x_{\rm c}<r_{\rm right}< r_2$.

Taking derivatives w.r.t.\ points $r_1$ and $r_2$ yields
\bea\label{lhs}
&&\!\!\!- \frac{\partial^2}{\partial {r_1}\partial r_2} \,\overline {\rme^{\int_x\lambda(x) S(x)}} \nn\\
&&\quad = \int_0^\infty {\rmd} S(x_c)\,  \rme^{\eta S(x_c)} {\cal P}(S(x_c),r_{\rm left},r_{\rm right}) \nn\\
&&\quad = P_{\ell}(r_2-r_1) \left<  \rme^{\eta S(x_c)} \right>_{\!r_1}^{\!r_2}  \nn\\
&&\quad = P_{\ell}(r_2-r_1)\left[1 + \eta \langle S(x_c) \rangle_{r_1}^{r_2} + \frac{\eta^2}2 \langle  S(x_c)^2 \rangle_{r_1}^{r_2} + ... \right] .\nn\\
\eea
Here $P_\ell(\ell)$ is the probability to have an avalanche with extension $\ell$, and   angular brackets $\langle... \rangle_{r_1}^{r_2}$ denote conditional averages given that the endpoints of the avalanches are at $r_1$ and $r_2$.

We now consider derivatives w.r.t.\ points $r_1$ and $r_2$ of the r.h.s.\ of \Eq{basic2}.  Using the expansion (\ref{31}) yields
\bea \label{rhs}
&&\!\!\!- \frac{\partial^2}{\partial {r_1}\partial r_2}  \rme^{\int_x   w(x) \tilde u(x) }  = - \rme^{\int \rmd x \,w(x) \tilde u_{0}(x)} \nn\\
&&\qquad \times \Big[\int \rmd x \,w(x) \frac{\partial^2 \tilde u_{0}(x)}{\partial r_1 \partial r_2}+ \eta \int \rmd x \,w(x) \frac{\partial^2 \tilde u_{1}(x)}{\partial r_1 \partial r_2} \nn\\
&&\qquad ~~+ \eta^2\int \rmd x \,w(x) \frac{\partial^2 \tilde u_{2}(x)}{\partial r_1 \partial r_2} + ...\Big]
\eea
Omitted terms indicated by $...$ are higher order in $w$. 
Comparing \Eqs{lhs} and \eq{rhs} yields for the probability to find an avalanche with extension $\ell$
\begin{align}
&P_{\ell}(\ell = r_2-r_1) = - \rme^{\int \rmd x \,w(x) \tilde u_{0}(x)}\int  \rmd x \,w(x) \frac{\partial^2 \tilde u_{0}(x)}{\partial r_1 \partial r_2} + ...
\end{align}
We now have to specify the driving. Following the discussion in section \ref{s:driving}, we either have to use uniform driving restricted to $[r_1+x_0, r_2-x_0]$, or  choose   random seeds $x_{\rm s}$  uniformly distributed  between $r_1$ and $r_2$.
Here we write formulas for the latter, choosing   $w(x) = w \delta (x-x_s) $. This yields 
\begin{align}
&P_{\ell}(\ell = r_2-r_1) = - w \int_{r_1}^{r_2} \rmd x_{\rm s}\,  \rme^{w \tilde u_{0}(x_{\rm s})} \frac{\partial^2 \tilde u_{0}(x_{\rm s})}{\partial r_1 \partial r_2} + ...
\end{align}
In the limit of small $w$ this becomes 
\begin{align}
&P_{\ell}(\ell = r_2-r_1) = - w \int_{r_1}^{r_2} \rmd x_{\rm s}\, \frac{\partial^2 \tilde u_{0}(x_{\rm s})}{\partial r_1 \partial r_2} + ...
\end{align}
Dropping the index s for the seed position, the final formulas for the observables of interest are 
\begin{align}
\label{44!}
&\!\!P_{\ell}(\ell = r_2-r_1) = - w \int_{r_1}^{r_2}  \rmd x \, \frac{\partial^2 \tilde u_{0}(x)}{\partial r_1 \partial r_2} \\
&\!\!P_{\ell}(\ell = r_2-r_1)  \left<   S(x_c) \right>_{\!r_1}^{\!r_2} = - w \int_{r_1}^{r_2}  \rmd x \, \frac{\partial^2 \tilde u_{1}(x)}{\partial r_1 \partial r_2} 
\label{45!}
 \\
&\!\!P_{\ell}(\ell = r_2-r_1) \half  \left< S(x_c)^2 \right>_{\!r_1}^{\!r_2}  = - w \int_{r_1}^{r_2}  \rmd x \,  \frac{\partial^2 \tilde u_{2}(x)}{\partial r_1 \partial r_2}
\label{46!}
\end{align}
The shape and its variance are thus given by the ratios of the above equation. 

The following calculations are structured as follows: 
In the next subsection, we give the instanton solution \eq{34} for extension $\ell=1$; more precisely $r_1=-1/2$, $r_2=1/2$. 

In a second step performed in section \ref{s:from Sbox to S}, we reconstruct the solution for general $r_1$ and $r_2$. This   allows us to vary as in \Eq{rhs} w.t.t.\ $r_1$ and $r_2$, thus selecting only those avalanches which touch the borders at $r_1$ and $r_2$. With the   normalization \eq{44} obtained from the probability to find an avalanche of extension $\ell$ performed in subsection \ref{s:PofL}, this allows us to give the normalized shape, and its fluctuations in subsection \ref{s:Results for the shape and its second moment}. 

\subsection{How to obtain the mean shape of all avalanches inside a box   of size 1, and its fluctuations}
We now solve \Eq{34} at $r_1=-1/2$, $r_2=1/2$. 
\begin{figure}
\fig{1}{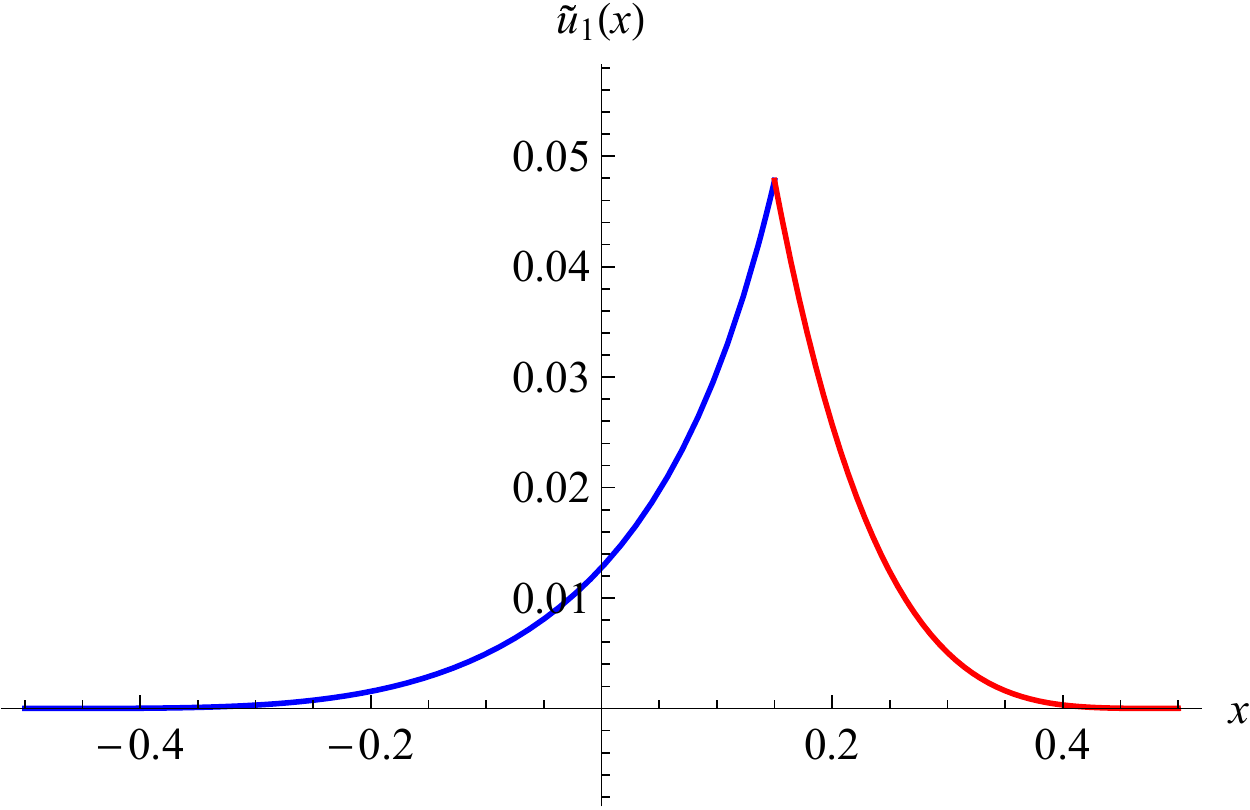}
\caption{The solutions (\ref{31}) at order $\eta$ for $x_{\rm c}=0.15$. Note that $ \tilde u_1(x)$ grows with a quartic power close to the boundary.}
\label{f-lambda-plot-order-1}
\end{figure}
One can write down differential equations to be solved by $  \tilde u_1(x)$ and $  \tilde u_2(x)$. 
There is, however,  a more elegant way to derive the perturbed instanton solution: To achieve this, we first realize that if $\tilde u(x)$ is a solution of $\tilde u ''(x)+  \tilde u(x)^2=0$, then 
$\tilde u_{\lambda,c}(x) := \lambda^2 \tilde u(\lambda x+c)$ is also a solution. 
We wish to construct  solutions which diverge at $x=\pm 1/2$, i.e.\ have extension 1, and which produce the additional term proportional to $\eta$ in Eq.~(\ref{30}). This can be achieved by separate solutions for the left branch, i.e.\ $-1/2<x<x_{\rm c}$, and the right branch $x_{\rm c}< x < 1/2$.  Using the symbol $f$ to indicate extension 1 as in \Eq{50}, we have
\bea
  f_{\lambda_{\rm L}}^{\rm L}(x) &:=& \lambda_{\rm L}^2 f\big(\lambda_{\rm L} (x+1/2)-1/2 \big) \\
x_{\lambda_{\rm L}}^{\rm L}(f) &=& \lambda_{\rm L}^{-1} x(\lambda_{\rm L}^{-2}f)+\frac12\left( \frac1{\lambda_{\rm L}} -1\right)\\ 
  f_{\lambda_{\rm R}}^{\rm R}(x) &:=& \lambda_{\rm R}^2 f\big({\lambda_{\rm R}} (x-1/2)+1/2 \big) \\
x_{\lambda_{\rm R}}^{\rm R}(f) &=& \lambda_{\rm R}^{-1} x( \lambda_{\rm R}^{-2}f)-\frac12\left( \frac1{\lambda_{\rm R}} -1\right)\ .~~
\eea
The two functions must coincide at $x_{\rm c}$, and   their slope must change by $\eta$;   more precisely 
\begin{align}
&  f_{\lambda_{\rm L}}^{\rm L}(x_{\rm c}) =  f_{\lambda_{\rm R}}^{\rm R}(x_{\rm c})\ ,\\
& \partial_{x } f_{\lambda_{\rm L}}^{\rm L}(x)\Big|_{x=x_{\rm c}} = \partial_{x}  f_{\lambda_{\rm R}}^{\rm R}(x)\Big|_{x=x_{\rm c}} +\eta \ . \label{40}
\end{align}
The second equation is written in a way to make clear that while $\lambda_{\rm L}$ and $\lambda_{\rm R}$ depend on $x_{\rm c}$, this dependence is not included in the derivatives of \Eq{40}.
We make the ansatz 
\bea
\lambda_{\rm L} &=& 1+a \eta + c \eta^2 \ ,\\
\lambda_{\rm R} &=& 1+ b \eta + d \eta^2\ .
\eea
Repeatedly using \Eqs{44}, \eq{45} and \eq{46} to eliminate higher derivatives, we find 
\bea
a&=&\frac{(2 x_{\rm c}-1) f'(x_{\rm c})+4 f(x_{\rm c})}{12
   \bar{\mathcal{E}}_1} \\
b&=& \frac{(2 x_{\rm c}+1) f'(x_{\rm c})+4 f(x_{\rm c})}{12
   \bar{\mathcal{E}}_1} \\
c&=& \frac1{288 \bar{\mathcal{E}}_1^2} \Big[16 f(x_{\rm c}) \left((1-3 x_{\rm c})
   f'(x_{\rm c})\right) \nn\\
   && +(2 x_{\rm c}-1) \Big(f'(x_{\rm c})
   \Big(\left(4 x_{\rm c}^2-1\right) f''(x_{\rm c}) \nn\\
   && +4
   \left(3 x_{\rm c}+1\right) f'(x_{\rm c})\Big)+24 \bar{\mathcal{E}}_1
   x_{\rm c}\Big)-96 f(x_{\rm c})^2 \Big]~~~~~\\
d&=& \frac1{288 \bar{\mathcal{E}}_1^2} \Big[ 16 f(x_{\rm c}) (-(3 x_{\rm c}+1)
   f'(x_{\rm c}))-96 f(x_{\rm c})^2  \nn\\
   && +(2 x_{\rm c}+1) \Big(f'(x_{\rm c}) \Big(\left(4 x_{\rm c}^2-1\right) f''(x_{\rm c}) \nn\\
   && +4 \left(3
   x_{\rm c}-1\right) f'(x_{\rm c})\Big)+24 \bar{\mathcal{E}}_1 x_{\rm c}\Big) \Big] ~~~~~
\ .
\eea
This gives 
\begin{widetext}
\bea
 f_{\lambda_{\rm L}}^{\rm L}(x) &=& f(x)+ \eta \frac{a}{2}     \Big[ (2 x+1) f'(x)+4 f(x)\Big]+\frac{ \eta ^2}{8}
   \Big[ 8  (a^2+2 c ) f(x)+ 4 (2 x+1)  (2 a^2+c ) f'(x)+a^2 (2 
   x+1)^2 f''(x)\Big]\nn\\
   && +O(\eta ^3) \\
    f_{\lambda_{\rm R}}^{\rm R}(x) &=& f(x)+\eta \frac{b}{2}   \Big[   (2 x-1) f'(x)+ 4 f(x) \Big] +\frac{ \eta ^2}{8}\Big[8
    (b^2+2 d ) f(x) + 4 (2 x-1)    (2 b^2+d ) f'(x)+ b^2 (2 x-1)^2 f''(x) \Big]\nn\\
   && +O(\eta ^3)
\ .
\eea
For illustration we plot on Fig.~\ref{f-lambda-plot-order-1} the order-$\eta$ solution  for $x_{\rm c}=0.15$.

We are finally   interested in uniformly distributed random seeds, i.e.\ we need to integrate these solutions over the driving point  $x$ inside the box, i.e.\ from $-1/2$ to $ 1/2$. To this purpose 
define 
\begin{align} \label{Fdef}
&\highlight{F(x):= 6  \zeta\! \left(\!x+\frac{1}{2};0,\frac{\Gamma \left(\frac{1}{3}\right)^{18}}{64 \pi
   ^6}\right) -F_0\ , \qquad F_0 = 6 \zeta\! \left(\! \frac{1}{2};0,\frac{\Gamma \left(\frac{1}{3}\right)^{18}}{64 \pi
   ^6}\right) \equiv 2 \pi \sqrt{3} \ , \qquad F'(x) = f(x) } \\
   & \highlight{F(0) = 0\ , \qquad F(x+1) = F(x) + 2 F_0    }\ .
\end{align}
Then, subtracting the solution at $\eta=0$ which is not needed (but whose integral is divergent), we obtain
\bea
\int_{-\frac12}^{x_{\rm c}} \rmd x\,  \left[ f_{\lambda_{\rm L}}^{\rm L}(x) -f(x)\right] &=&   \eta a \Big[ \frac{1}{2}   (2 x_{\rm c}+1) f(x_{\rm c})+ F_0+F(x_{\rm c})\Big] \nn\\
&& 
+\eta^2 \Big[ \frac{1}{8} a^2 (2 x_{\rm c}+1)^2 f'(x_{\rm c})+\frac{1}{2}
   (a^2+c) (2 x_{\rm c}+1) f(x_{\rm c})+c \big(F(x_{\rm c})+F_0\big) \Big] \nn\\
\int_{x_{\rm c}}^{\frac12}  \rmd x\, \left[  f_{\lambda_{\rm R}}^{\rm R}(x) -f(x)\right] &=&   \eta  b \Big[ \frac{1}{2}   (1-2 x_{\rm c}) f(x_{\rm c})+F_0-F(x_{\rm c})  \Big] \nn\\
&& 
+\eta^2 \Big[-\frac{1}{2} (b^2+d) (2 x_{\rm c}-1)
   f(x_{\rm c})-\frac{1}{8} b^2 (2 x_{\rm c}-1)^2
   f'(x_{\rm c})+ d \big(F_0-F(x_{\rm c}) \big)    \Big]
\ .
\eea
This yields the (unnormalized) expectation, given that the interface has not moved at points $\pm 1/2$:  
\begin{align}\label{86}
&  \left< \rme^{\eta S  (x_{\rm c})}-1\right>= w\int_{-\frac12}^{x_{\rm c}} \rmd x\,  \left[ f_{\lambda_{\rm L}}^{\rm L}(x) -f(x)\right] +w \int_{x_{\rm c}}^{\frac12}  \rmd x\,  \left[ f_{\lambda_{\rm R}}^{\rm R}(x) -f(x)\right] \nn\\
& =   w \eta \Big[ \frac{1}{2} f\left(x_{\rm c}\right) \Big(2 (a-b) x_{\rm c}+a+b\Big)+(a-b)F(x_{\rm c})+(a+b)F_0
    \Big] \nn\\
&
~~~~+w \frac{\eta^2}8 \bigg[ 4 f (x_{\rm c}) \Big(2 x_{\rm c}
   (a^2-b^2+c-d)+a^2+b^2+c+d\Big)+\Big(2 (a-b) x_{\rm c}+ a+b\Big)
   \Big(2 (a+b) x_{\rm c}+a-b\Big) f'(x_{\rm c}) \nn\\
   &~~~~ \qquad ~~~ +8 (c-d)  F(x_{\rm c}) +8(c+d)F_0\big) \bigg] + ...  \nn\\
   & = w \eta\, \frac{2 f(x_{\rm c}) \big[f(x_{\rm c})+2
   F_0\big]-\big[F(x_{\rm c})-2 F_0 x_{\rm c}\big]
   f'(x_{\rm c})}{6 \bar{\mathcal{E}}_1} \nn\\
   & ~~~~+ w \frac{\eta^2}{144 \bar{\mathcal{E}}_1^2}\bigg[  -4 x_{\rm c} F(x_{\rm c}) \Big(6
   \bar{\mathcal{E}}_1+f'(x_{\rm c})^2\Big)+4 F_0 \Big(12
   \bar{\mathcal{E}}_1 x_{\rm c}^2+(6 x_{\rm c}^2-1)
   f'(x_{\rm c})^2 \Big)+f(x_{\rm c})^2 \Big(8 x_{\rm c}
   f'(x_{\rm c})-96 F_0\Big) \nn\\
   &~~~~+f(x_{\rm c}) f'(x_{\rm c})
   \Big(3 (4 x_{\rm c}^2-1) f'(x_{\rm c})+16
   F(x_{\rm c})-48 F_0 x_{\rm c}\Big)+(4 x_{\rm c}^2-1) \Big(2
   F_0 x_{\rm c}-F(x_{\rm c})\Big) f'(x_{\rm c})
   f''(x_{\rm c})-32 f(x_{\rm c})^3\bigg]\nn\\
    &~~~~+ ...\nn\\
   &=: w \eta S_{\rm box} ^{\ell=1}(x_{\rm c}) + w \frac{\eta^2}{2} S_{\rm box} ^{2,\ell=1}(x_{\rm c}) + ...\ .
\end{align}
\end{widetext}
We have termed these expressions $S_{\rm box} ^{\ell=1}(x_{\rm c})$ and $S_{\rm box} ^{2,\ell=1}(x_{\rm c})$. We recall that this is not yet the sought-for avalanche shape, and fluctuations. Rather, it is the expectation of the size $S(x)$ inside a box of size  $1$, given that the avalanche does not touch any of the two boundaries $x=\pm 1/2$. We will have to vary the boundary points in order to extract the shape $\left< S(x) \right>$ of avalanches which vanish at the boundary points, but not before.  This is the objective of the next section. 

For later reference, we note 
\begin{align}\label{87}
&\int_{-\frac12}^{\frac12} \rmd x\,  S_{\rm box}(x) =\nn\\
   & =  -\frac{3 f'(x)+f(x) F(x)-2 F_0 (x f(x)+F(x))}{6 \bar{\mathcal{E}}_1}\Bigg|_{-\frac12}^{\frac12}\nn\\
   &= \frac{2 F_0^2}{3 \bar{\mathcal{E}}_1} = \frac{256 \pi ^8}{9 \Gamma
    (\frac{1}{3} )^{18}} = 0.00534401\ , \\
   \label{87-2}
& \int_{-\frac12}^{\frac12} \rmd x_{\rm c} \, S_{\rm box} ^{2,\ell=1}(x_{\rm c})  =   2.3030 \times 10^{-6}  \ .
\end{align}
\vspace{3.5cm}

\section{From $S_{\rm box}(x)$ to the shape $S(x)$: Scaling arguments, etc.}
\label{s:from Sbox to S}
\subsection{The probability to find an avalanche of extension $\ell$, and probability for   seed  position}
\label{s:PofL}
The probability to have an avalanche of size $\ell$ is according to Eqs.~(\ref{8})  and \eq{44} to leading order in $w$ given by
\begin{align}
&\highlight{P_{\ell }(\ell)}
= - w \int_{r_1}^{r_2}  \rmd x \, \frac{\partial^2 \tilde u_{0}(x)}{\partial r_1 \partial r_2}  \nn\\
 &=  \frac w{\ell^3} \int_{-\frac12}^{\frac12}\rmd x \, \frac{   (4 x^2-1) f''(x)+24 (x f'(x)+f(x))}{4 } + ...   \nn\\
 & \simeq \highlight{ 4 F_0 \frac w{\ell^3} }+ ... \equiv \highlight{ 8 \pi \sqrt{3} \frac w{\ell^3} }+ ...\ ,\label{10}
\end{align}
with $F_0$ defined in \Eq{Fdef}.  

It is interesting to note that the integrand in \Eq{10} gives the probability to have the seed at position $x$. More precisely, the probability that the seed was at position $x$ inside an avalanche extending from $-1/2$ to $1/2$ is  
\be
P^{\rm seed}_{\ell=1}(x) = \frac{   (4 x^2-1) f''(x)+24 [x f'(x)+f(x)]}{32 \pi \sqrt{3} }\ .
\ee
This function starts with a cubic power at the boundary. We give a series expansion below in \Eq{PseedSeries}.

\subsection{Basic scaling relations, and consequences}
In general, the size of an avalanche scales as $S (\ell)\sim \ell^{{d+\zeta}}$. For the BFM, the latter reduces to 
\be\label{53}
S (\ell)\sim \ell^{4} \ .
\ee
The proportionality  constant is calculated in Eq.~(\ref{62}) below. 

Let us now solve the instanton equation (\ref{34}) with source (\ref{30})  for arbitrary $r_1$ and $r_2$. This can be  achieved by observing that, as a function of $|r_2-r_1|$, 
\bea
\tilde u''(x) &\sim& \tilde u(x)^2 \sim \frac1{|r_2-r_1|^4} \nn\\
&\sim& \eta \delta (x-x_c) \equiv \frac{\eta}{|r_2-r_1|} \delta \left(\frac{x-x_c}{|r_2-r_1|}\right)\ . ~~~
\eea
Thus $\eta \sim |r_2-r_1|^{-3}$, and  
\begin{align}\label{60}
&  \tilde u^{x_{\rm c}}_{1,r_2,r_1}(x) =   |r_2-r_1|\, \tilde u^{\frac{x_{\rm c}-(r_1+r_2)/2}{r_2-r_1}}_{1,r_{1/2}=\mp \half} \Big(\textstyle{ \frac{x-(r_1+r_2)/2}{r_2-r_1}}\Big)\ ,   \\
&\!\! \Rightarrow  S_{\rm box}^{r_1,r_2}(x_{\rm c}) = \int_{r_1}^{r_2} \rmd x\, \tilde u^{x_{\rm c}}_{1,r_2,r_1}(x)  \nn\\
&
\hphantom{\Rightarrow  S_{\rm box}^{r_1,r_2}(x_{\rm c})} =
  |r_2-r_1|^2 \,S_{\rm box}^{r_2-r_1=1} \Big(\textstyle{ \frac{x_{\rm c}-(r_1+r_2)/2}{r_2-r_1}}\Big)\label{61}
\ .
\end{align}
This is consistent with the dimension of an avalanche $S$ per length $\ell$, i.e.\ $S/ \ell \sim \ell^3$.

Now, the (unnormalized) shape of an avalanche of extension $\ell$ is according to \Eq{45!} obtained as 
\bea
S_{\ell = r_2-r_1}(x) &=& -\partial_{r_{2}}\partial_{r_{1}} S^{r_{2},r_{1}}_{\rm box}(x) \ .
\eea
Using Eq.~(\ref{60}), this yields
\bea\label{66}
\highlight{S_{\ell=1}(x) }&=& -\partial_{r_{2}}\partial_{r_{1}}\Big[ |r_2{-}r_1|^2 S_{\rm box}^{\ell=1}\Big(\textstyle{ \frac{x-(r_1+r_2)/2}{r_2-r_1}}\Big) \Big]_{r_{1}= - \frac12}^{r_2=\frac12} \nn\\
&=&\highlight{\Big[ 2 -2 x \,\partial_x +  \Big(x^2-\frac 14\Big)\partial_x^2\Big] S_{\rm box}^{\ell=1}(x)}\ .
\eea
We note that this function grows cubicly at the boundary, consistent with our scaling argument (\ref{gofx}). 
To achieve this, the factor of $|r_2{-}r_1|^2$ in Eq.~(\ref{66}) is crucial: Were the exponent larger than $2$, then the growth would be linear. Were it smaller, the  function (\ref{66}) would become negative. 

Integrating by parts we obtain using \Eq{87} 
\be\label{64}
\int_{-\frac12}^{\frac12}\rmd x\, S_{\ell=1}(x) = 6 \int_{-\frac12}^{\frac12} \rmd x_{\rm c}\,S_{{\rm box}}^{\ell=1}(x_{\rm c}) = \frac{4 F_0^2}{  \bar{\mathcal{E}}_1}
\ .
\ee

Similarly, we find for the order-$\eta^2$ term
\bea
\highlight{S^{2}_{\ell=1}(x)} &=& -\partial_{r_{2}}\partial_{r_{1}}\Big[ |r_2{-}r_1|^5 S_{\rm box}^{2,\ell=1}\Big(\textstyle{ \frac{x-(r_1+r_2)/2}{r_2-r_1}}\Big) \Big]_{r_{1}= - \frac12}^{r_2=\frac12} \nn\\
&=&\highlight{ \Big[ 20 -8 x \,\partial_x +  \Big(x^2-\frac 14\Big)\partial_x^2\Big] S_{\rm box}^{2,\ell=1}(x) }
\ .
\eea
This implies
\be
\int_{-\frac12}^{\frac12}\rmd x\, S^{2,\ell=1}(x) = 30\int_{-\frac12}^{\frac12} \rmd x_{\rm c}\,S_{{\rm box}}^{2,\ell=1}(x_{\rm c})
\ .
\ee
Note that according to \Eqs{44!}-\eq{46!} $S_{\ell}(x) $ and $S_{\ell}^2(x) $ are not yet properly normalized to give the expectation of the shape of an avalanche. 
For this purpose, let us define with the help of Eq.~(\ref{10})
\bea
  \left< {S (x) }\right>_{\ell} &:=& \frac{ w {S_{\ell}(x) }  }{P_{\rm aval}(\ell)} = \frac{S_{\ell=1}(x/\ell)}{4 F_0}\ell^3\ , \\
 \left< {S ^2(x) }\right>_{\ell} &:=& \frac{ w {S_{\ell}^2(x) }  }{P_{\rm aval}(\ell)} = \frac{S^2_{\ell=1}(x/\ell)}{4 F_0}\ell^6
\ .
\eea
These functions give the shape of an avalanche {\em given that the avalanche extends from $-\half $ to $\half$}, as well as its fluctuations, {\em including the amplitude}. 

For the total size $\left< S\right>_\ell = \int_{-\frac \ell 2}^{\frac  \ell 2} \rmd x\, \left< S(x)\right>_{\ell}$ and the integral of the second moment $\left< S^2(x)\right>_\ell$ we find
\begin{align}\label{62}
&{\int_{-\frac\ell 2}^{\frac{\ell }{2}} \rmd x \left< {S (x) }\right>_{\ell} } = \frac{F_0}{\bar {\cal E}_1} \ell^4 = 0.000736576\, \ell^4\ ,
\\
&{\int_{-\frac\ell 2}^{\frac{\ell }{2}} \rmd x  \left< {S ^2(x) }\right>_{\ell} }  =5.29044 \times 10^{-8}\,
\ell^7\ .
\end{align}

\subsection{Results for the shape and its second moment}
\label{s:Results for the shape and its second moment}
\begin{figure*}
\fig{1}{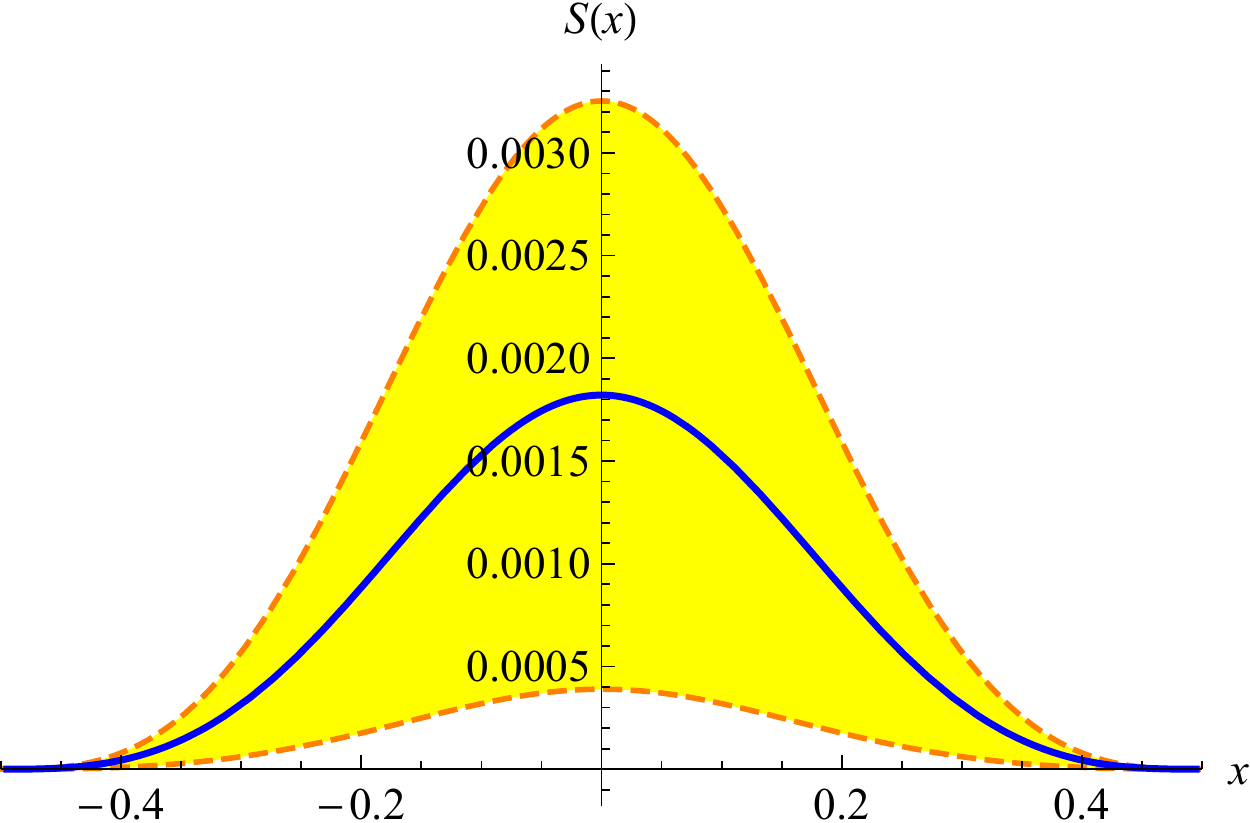}~~~~~\fig{1}{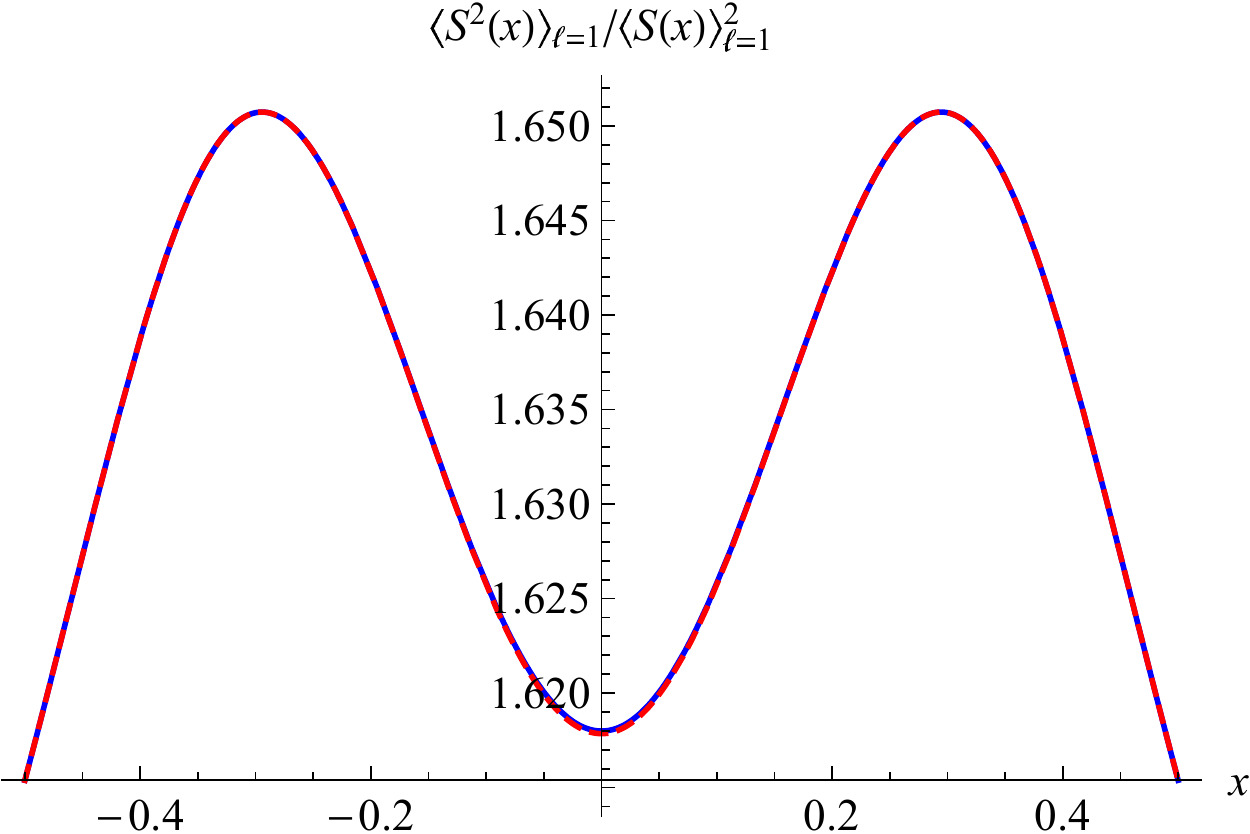}
\caption{Left: The spatial shape $\left<S(x)\right>_{\ell=1}$ of an avalanche conditioned conditioned to have  size $\ell=1$ (blue, solid line).  The dashed curves represent $\left<S(x)\right>_{\ell=1}\pm \sqrt{\left< {S_{\ell}^2(x) }\right>_{\ell=1}^{\rm c} }$. Right: The ratio $\left< {S_{\ell}^2(x) }\right>_{\ell=1}/\left< {S_{\ell}(x) }\right>_{\ell=1}^2$, which has spatial average (integeral) $1.63523384$; blue from an interpolating function, red dashed from series expansion with the 16 leading terms; note that below in \Eq{expS2overexpSsquared} only the leading 8 are given.}
\label{f:S+S2}
\end{figure*}
We give explicit formulas for $\left<S(x)\right>_{\ell=1} $, and $\left<S^2(x)\right>_{\ell=1} $ below. They are plotted on Fig.~\ref{f:S+S2}. We did not succeed in finding much simpler expressions. While especially the expression for the second moment $\left<S^2(x)\right>_{\ell=1} $ is lengthy, its ratio with the squared  first moment is almost constant, given by 
\be
\frac{\left<S^2(x)\right>_{\ell=1}}{\left<S(x)\right>_{\ell=1}^2} \approx 1.635 \pm 0.02\ .
\ee
This can be seen on Fig.~\ref{f:S+S2}.
The explicit formulas are  
\begin{widetext}
\begin{align}\label{expS(x)}
&\left< {S (x) }\right>_{\ell=1} = \frac{1}{48 F_0 \bar{\mathcal{E}}_1}\bigg[ 3 (4 x^2-1) \bar{\mathcal{E}}_1-\Big(f(x) (-4 x^2 F(x)+F(x)+12 x)+2 F_0 x ((4 x^2-1) f(x)+8)+4 F(x)\Big) f'(x) \nn\\
& \qquad\qquad\qquad\qquad\quad +3 (4 x^2-1) f'(x)^2+4 f(x) (f(x) (-x (F(x)+2 F_0 x)+F_0+2)+4 F_0)\bigg]\ , 
\end{align}
\begin{align}
&\left< {S^2 (x) }\right>_{\ell=1} =\frac{1}{576 F_0 \bar{\mathcal{E}}_1^2}\bigg[ F(x) (-1344 x \bar{\mathcal{E}}_1+(1-4 x^2)^2 f'(x)^3-736 x 
f'(x)^2) \nn\\
& +2 F_0 \Big(192 (3 x^2 
\bar{\mathcal{E}}_1+\bar{\mathcal{E}}_1)+(-(1-4 x^2)^2 x f'(x)+384 
x^2+88) f'(x)^2\Big) \nn\\
& +f(x) \Big(24 (12 x^2+1) \bar{\mathcal{E}}_1-4 ((8 x^3-2 
x) F(x)+F_0 (32 x^4-20 x^2+3)+72 x^2+18) f'(x)^2+640 (F(x)+3 F_0 x) 
f'(x)\Big)\nn\\
&-6 x (4 x^2-1) f'(x) \Big(14 \bar{\mathcal{E}}_1+9 f'(x)^2\Big)+4 
f(x)^3 \Big((4 x^2-1) (-4 x^2 F(x)+2 F_0 (4 x^2-1) x+F(x)-2 x) 
f'(x)-320\Big)\nn\\
& +f(x)^2 \Big(f'(x) (-11 (1-4 x^2)^2 f'(x)-4 (52 x^2+3) 
F(x)+1472 x)+16 F_0 (x (52 x^2-5) f'(x)-240)\Big)\nn\\
& +f(x)^4 \Big(4 x (4 x^2-1) 
F(x)+F_0 (64 x^4-40 x^2+6)+8 (28 x^2-3)\Big)+2 (1-4 x^2)^2 f(x)^5\bigg]\ .
\end{align}
While these expressions are cumbersome, one can work with a converging Taylor series. 
An expansion in $(\frac12-x)(\frac12+x)$ respecting the Taylor expansion at the boundary is 
\bea\label{expS}
\left< {S (x) }\right>_{\ell=1} &=&  
\frac{1}{21} \left(\frac{1}{4}-x^2\right)^{\!3}
+\frac{3}{28} \left(\frac{1}{4}-x^2\right)^{\!4}
+\frac{2}{7} \left(\frac{1}{4}-x^2\right)^{\!5} 
+\frac{5}{6} \left(\frac{1}{4}-x^2\right)^{\!6}
+\left(\frac{18}{7}-\frac{\bar{\mathcal{E}}_1}{1540 F_0}\right)  \left(\frac{1}{4}-x^2\right)^{\!7}  \nn\\
&& +\left(\frac{1}{4}-x^2\right)^{\!8} \left(\frac{33}{4}-\frac{7 \bar{\mathcal{E}}_1}{1760 F_0}\right)
+\left(\frac{1}{4}-x^2\right)^{\!9} \left(-\frac{\bar{\mathcal{E}}_1}{55 F_0}
+\frac{5 \bar{\mathcal{E}}_1}{34398}+\frac{572}{21}\right)
\nn\\
&&
+\left(\frac{1}{4}-x^2\right)^{10} \left(\frac{3\left(\bar{\mathcal{E}}_1+78078\right)}{2548}-\frac{3 \bar{\mathcal{E}}_1}{40 F_0}\right)
+ ...  \\
\label{expS2}
\left< {S (x)^2 }\right>_{\ell=1} &=& \frac{1}{273} \left(\frac{1}{4}-x^2\right)^{\!6}  + \frac{5}{294} \left(\frac{1}{4}-x^2\right)^{\!7} + \frac{503}{7644} \left(\frac{1}{4}-x^2\right)^{\!8} +\frac{309}{1274} \left(\frac{1}{4}-x^2\right)^{\!9}\nn\\
&& +
\left( \frac{561}{637}-\frac{529 \bar{\mathcal{E}}_1}{3898440 F_0} \right) \left(\frac{1}{4}-x^2\right)^{\!10}
+ \left( \frac{937}{294}-\frac{8641 \bar{\mathcal{E}}_1}{7796880 F_0}\right)  \left(\frac{1}{4}-x^2\right)^{\!11}
\nn\\
&& 
+\left( -\frac{531133 \bar{\mathcal{E}}_1}{85765680 F_0}+\frac{\bar{\mathcal{E}}_1}{25137}+\frac{485}{42}\right)  \left(\frac{1}{4}-x^2\right)^{\!12} + ... 
\eea
\bea
\frac{\left< {S (x)^2 }\right>_{\ell=1}}{\left< {S (x) }\right>^2_{\ell=1}} &=& \frac{21}{13} +\frac{3}{13} \left(\frac{1}{4}-x^2\right)+\frac{87}{208} \left(\frac{1}{4}-x^2\right)^{\!2}+\frac{411}{416} \left(\frac{1}{4}-x^2\right)^{\!3}+ 
\left(\frac{1}{4}-x^2\right)^{\!4} \left(\frac{8877}{3328}-\frac{307 \bar{\mathcal{E}}_1}{19448 F_0}\right)\nn\\
&& + \left(\frac{1623}{208}-\frac{127 \bar{\mathcal{E}}_1}{2992 F_0}\right) \left(\frac{1}{4}-x^2\right)^{\!5} 
+ \left(-\frac{31591 \bar{\mathcal{E}}_1}{311168 F_0}+\frac{74 \bar{\mathcal{E}}_1}{9633}+\frac{1281987}{53248}\right)\left(\frac{1}{4}-x^2\right)^{\!6}
\nn\\
&& + \left(-\frac{732863 \bar{\mathcal{E}}_1}{3111680 F_0}+\frac{5543 \bar{\mathcal{E}}_1}{134862}+\frac{8216901}{106496}\right) \left(\frac{1}{4}-x^2\right)^{\!7}+ ...
\label{expS2overexpSsquared}
\ .
\eea
For completeness, we also give a series expansion for $P^{\rm seed}_{\ell=1}(x)$,
\bea \label{PseedSeries}
P^{\rm seed}_{\ell=1}(x) &=& \frac{\bar{\mathcal{E}}_1}{8 \sqrt{3} \pi } \bigg[ \frac{2}{7}
   \left(\frac{1}{4}-x^2\right)^{\!3} +\frac{5}{14}
   \left(\frac{1}{4}-x^2\right)^{\!4}+ \frac{4}{7} \left(\frac{1}{4}-x^2\right)^{\!5}
   +\left(\frac{1}{4}-x^2\right)^{\!6}
   +\frac{12}{7} \left(\frac{1}{4}-x^2\right)^{\!7}
   +\frac{33}{14} \left(\frac{1}{4}-x^2\right)^{\!8} \nn\\
   && +\frac{5  \bar{\mathcal{E}}_1}{22932}\left(\frac{1}{4}-x^2\right)^{\!9}
   + \frac{11 
   \left(\bar{\mathcal{E}}_1-12168\right)}{6552}\left(\frac{1}{4}-x^2\right)^{\!10}
   +\frac{ 205
   \bar{\mathcal{E}}_1-2895984}{22932} \left(\frac{1}{4}-x^2\right)^{\!11}+ ...~
   \bigg]\ .
\eea

\end{widetext}

\label{s:numerics}

\fboxsep0mm
\newcommand{\kbox}[1]{{\!\!{#1}}}
\setlength{\unitlength}{1.02cm}
\begin{figure}
\kbox{\begin{picture}(8.4,5.7)
\put(0.5,0){\cfig{8\unitlength}{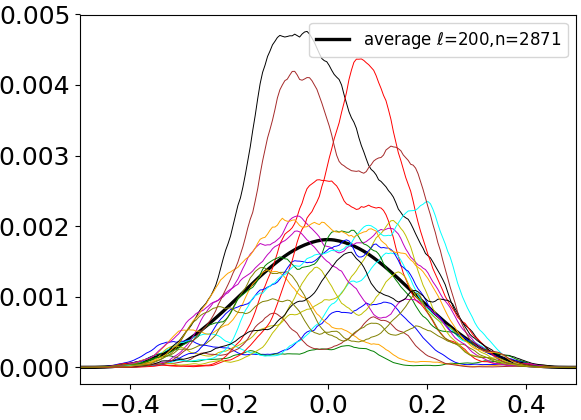}}
\put(0,5){\rotatebox{90}{$S(x)$}}
\put(8.2,0.1){$x$}
\end{picture}}
\caption{20 avalanches with extension $\ell=200$, rescaled to $\ell=1$.  $n=2871$ is the number of samples used for the average.}
\label{f:samples}
\bigskip
\kbox{\begin{picture}(8.4,6.2)
\put(0.5,0.2){\cfig{8\unitlength}{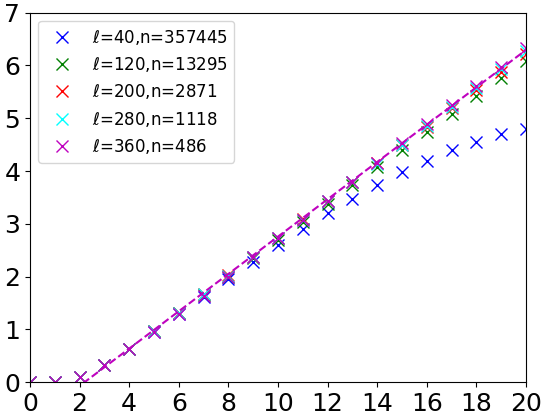}}
\put(0,4.25){\rotatebox{90}{$\sqrt[3]{\left<{S(x-r_1)}\right>}$}}
\put(8.2,0){$x$}
\end{picture}}
\caption{The function $\sqrt[3]{\left<{S(x-r_1)}\right> }$ at given $\ell$ becomes linear starting at about the third non-vanishing point. This leads to an effective offset of $2$ for the size. An extrapolation is shown for  $\ell=360$.}
\label{f:collapseSend}
\bigskip
\kbox{\begin{picture}(8.4,6.2)
\put(0.5,0.3){\cfig{8\unitlength}{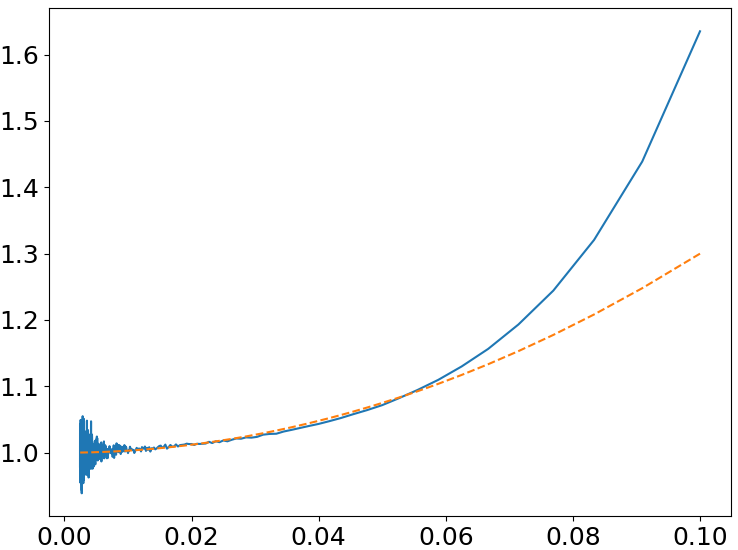}}
\put(0,4.3){\rotatebox{90}{$\langle S \rangle/\langle S \rangle_{\rm theory}$}}
\put(8.,0){$1/\ell$}
\put(6.8,2.5){$1+30/\ell^2$}
\end{picture}}
\caption{The mesaured avalanche size $\left<S\right>$, as a function of $\ell$, divided by the theory prediction from Eq.~(\ref{62}). The dashed orange line is the estimated finite-size correction $1+30/\ell^2$.}
\label{f:Stot}
\end{figure}\begin{figure} 
\kbox{\begin{picture}(8.4,5.9)
\put(0.5,0){\cfig{8\unitlength}{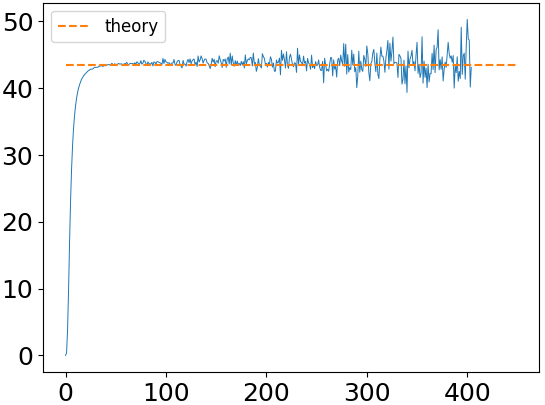}}
\put(0,5){\rotatebox{90}{$P(\ell) \ell^3 $}}
\put(8.2,0.05){$\ell$}
\end{picture}}
\caption{The rescaled distribution of extensions $P(\ell)\ell^{3}$ as a function of $\ell$. }
\label{f:Pofl}
\bigskip\kbox{\begin{picture}(8.4,6.05)
\put(0.5,0.){\cfig{8\unitlength}{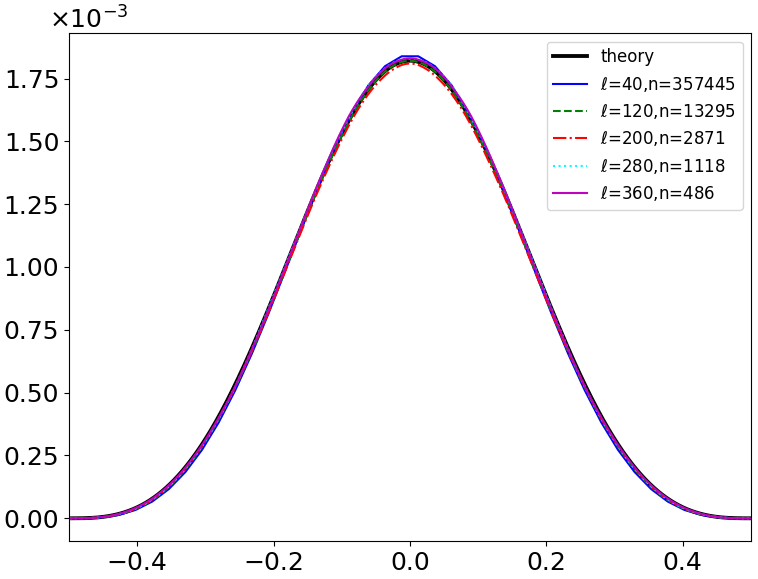}}
\put(0,4.9){\rotatebox{90}{$\left<{S(x)}\right>$}}
\put(8.2,0.1){$x$}
\end{picture}}
\caption{The shape $\left< S(x)\right>\equiv \left< S(x/\ell)\right>_{\ell } /\ell^3$ averaged for all avalanches with a given $\ell$  between $40$ and $360$. To reduce  statistical errors, we have symmetrised this function. The convergence is very good; this can best be seen on the error plot of Fig.~\ref{f:differences} (left).}
\label{f:StimesL3}
\bigskip\kbox{\begin{picture}(8.4,5.8)
\put(0.5,0.){\cfig{8\unitlength}{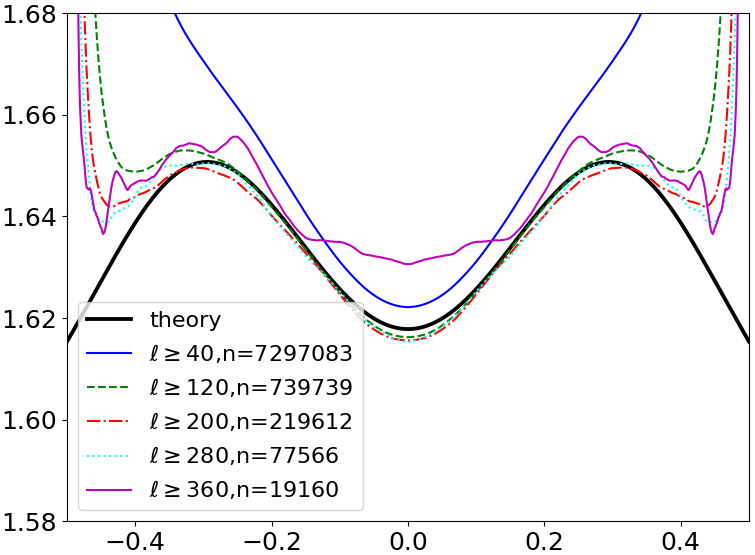}}
\put(0,3.4){\rotatebox{90}{$ \left< {S^2(x) }\right>/\left< {S(x) }\right>^2 $}}
\put(8.2,0.1){$x$}
\end{picture}}
\caption{The symmetrized ratio $\left< {S^2(x) }\right>/\left< {S(x) }\right>^2 \equiv \left< {S^2(x/\ell) }\right>_\ell/\left< {S(x/\ell) }\right>_\ell^2$, averaged for $\ell\ge \ell_{0}$. Convergence to the theoretical prediction in the boundary region is slow.}
\label{f:ratio}
\end{figure}
\setlength{\unitlength}{1.025cm}
\begin{figure*}\kbox{\begin{picture}(8.5,6.)
\put(0.5,0){\cfig{8\unitlength}{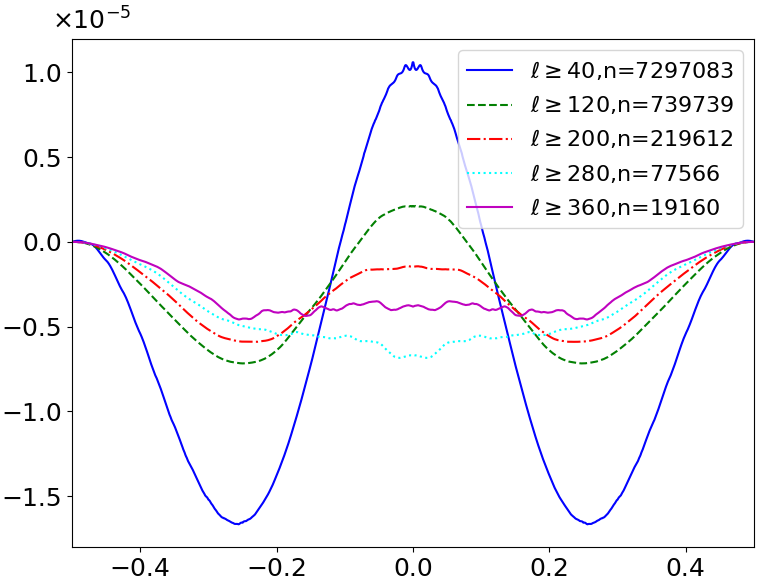}}
\put(0,3.){\rotatebox{90}{$  \left< S(x)\right>-\left< S(x)\right>_{\rm theory} $}}
\put(8.3,0.1){$x$}
\end{picture}}~~~~~\setlength{\unitlength}{1.043cm}\kbox{\begin{picture}(8.2,5.7)
\put(0.3,0.){\cfig{8\unitlength}{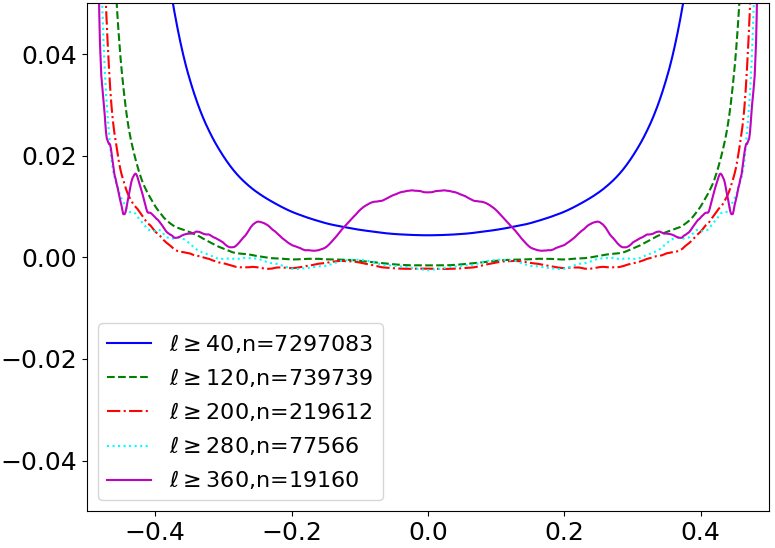}}
\put(0,2.2){\rotatebox{90}{$ \left< {S^2(x) }\right>/\left< {S(x) }\right>^2-\mbox{theory} $}}
\put(8.0,0.1){$x$}
\end{picture}}
\caption{Left: Error for the estimation of $\left< S(x)\right>  \equiv \left< S(x\ell)\right>_{{\ell}}/\ell^{3} $ minus its theoretical prediction, averaged over all avalanches with extension $\ell$ larger than a cutoff as given in the legend. We see that the systematic error decreases for increasing size, while the statistical error grows. The optimum value of $\ell$ is around $\ell=280$, with a relative error of about $ 3\times 10^{-3}$ in the center region. Right: ibid for the ratio $\left< {S^2(x) }\right>/\left< {S(x) }\right>^2 \equiv \left< {S^2(x/\ell) }\right>_\ell/\left< {S(x/\ell) }\right>_\ell^2$. Convergence in the boundary region $x\to \pm \half$ is slow, i.e.\ finite-size effects are important there.}
\label{f:differences}
\end{figure*}

\section{Numerical validation}
We verified our findings with large-scale numerical simulations. To this aim, we consider the equation of motion discretized in space, started with a kick of size 1, 
\begin{align}
&\partial_t \dot u_i(t) = \dot u_{i+1}(t)+\dot u_{i-1}(t)-2 \dot u_i(t) + \sqrt{\dot u_i(t)}\xi(t)\\
&\left< \xi(t) \xi(t')\right> = \delta (t-t')\\
&\dot u_0(0) =  \delta_{i,0}
\end{align}
Since we work in the Brownian force model, these equations {\em do not depend on the shape of the interface before the avalanche},  
and  one can always start from a flat interface. This would not be the case for finite-ranged disorder. 
For the same reason, we can choose to put the seed at zero, and to not  change the seed-position between avalanches.  

One further has to discretize in time, using a step-size $\delta t$. A naive implementation would lead to a factor of $\sqrt{\delta t}$ in front of the noise term. Thus the limit of $\delta t\to 0$ is difficult to take. Here we use an algorithm proposed in \cite{DornicChateMunoz2005}, and further developed for the problem at hand in \cite{DelormeLeDoussalWiese2016}. 
The idea is to use the conditional probability   $P(\dot u_i(t+\delta t)|\dot u_i(t) ,\dot u_{i_\pm 1}(t))$, where $\dot u_{i_\pm 1}(t)$ are assumed to remain fixed. From this probability, which is a Bessel function, is then drawn $\dot u_i(t+\delta t)$. Sampling of the Bessel function is achieved by its clever decomposition into a sum of Poisson times Gamma functions, for which efficient algorithms are available. This algorithm   scales linear with the time-discretization $\delta t$. 
It is   explained in details in Ref.~\cite{DelormeLeDoussalWiese2016},  appendix H.  

We run our simultions for a system of size $410$, time step $\delta t=0.01$,  producing a total of  $526929535$ avalanches. Since $P(\ell)\sim 1/\ell^3$, most avalanches have a small extension, and the statistics for them will be good. On the other hand, small avalanches have  important finite-size corrections, thus are not in the scaling limit.  In the following, we will show all our data, reminding of these two respective short-comings. 

Let us start by showing 20 avalanches of extension 100, see Fig.~\ref{f:samples}. One sees that that there are substantial fluctuations in the shape, roughly consistent with the theoretically expected domain plotted in Fig.~\ref{f:S+S2}. 

Let us next study the shape of the {\em discretized} avalanches close to the boundary. To this aim we plot on Fig.~\ref{f:collapseSend} the mean shape of all avalanches with a given size, taken to the power $1/3$. One sees that for a given point $i$ from the boundary, these curves converge against a limit when increasing $\ell$. This confirms our {\em boundary-shape conjecture} made in the introduction. 
Second, we see that the shape taken to the power $1/3$ converges against a straight line with slope $\sqrt[3]{1/21}$, as predicted, see Eqs.~(\ref{B}) and ~(\ref{expS}). However, there is a non-vanishing boundary-layer length $\ell_{\rm B}$, s.t.\ 
\be
\left< S(x-r_1) \right> \simeq \frac1{21}  (x-r_1- \ell_{\rm B})^3 + ...
\ .
\ee
Our extrapolations on Fig.~\ref{f:collapseSend} show that 
\be
\ell_{\rm B}   \approx 2\ .
\ee
In order to faster converge to the field-theoretic limit, we define   the total extension $\ell$ of an avalanche to be
\be\label{ldef}
\ell:= \ell_{\rm discretized} -2 \ell_{\rm B}\ ,
\ee
where $\ell_{\rm discretized}$ is the number of  points which advanced in an  avalanche. This definition can be interpreted such that the avalanche extends to the middle between the first non-moving point and the first moving one. As such, it contains some arbitrariness. The choice is motivated as follows: A good test object is the total size $\left<S\right>_\ell$ of an avalanche of extension $\ell$, which we know from Eq.\ (\ref{62})  to be
\be
\left< S\right>_\ell=  0.000736576\, \ell^4 .
    \label{87-bis}
\ee
Fig.~\ref{f:Stot} confirms this; it also shows that the approach to this limit has  finite-size corrections, which we estimate as 
\be
\left< S\right> \simeq  0.000736576\, \ell^4  \left[ 1+ \frac{30}{ \ell^{2}}+ {\cal O}(\ell^{-3})\right]
\ .
\ee
It is important to note that the curve enters with slope 0 into the asymptotic value at $\ell=\infty$, which is the best one can achieve with a linear shift in $\ell$. This makes us confident that our definition (\ref{ldef}) is indeed optimal.

We also note that the size of the kick puts an effective small-scale cutoff on the extension of avalanches. 
This can be seen on Fig.~\ref{f:Pofl}: One first verifies that the amplitude conforms to Eq.~(\ref{10}). Demanding that $\int_{\ell_{\rm c} }^{\infty} P(\ell) \rmd \ell = 1$ yields 
\be
\ell_{\rm c } \approx 2 \sqrt{\sqrt3 \pi w } = 4.665
\ee
We now come to a check of the shape itself. To this aim, we plot in Fig.~\ref{f:StimesL3} the mean shape of our avalanches,  rescaled to $\ell=1$. We see that these curves converge rather nicely to the predicted universal shape (\ref{expS(x)}), even for relatively small sizes. 

We then turn to the fluctuations. On Fig.~\ref{f:ratio} we plot the ratio  $\left< {S_{\ell}^2(x) }\right>_{\ell=1}/\left< {S_{\ell}(x) }\right>_{\ell=1}^2$. A glance at the right of figure \ref{f:S+S2} shows that it is almost constant, equal to $ 1.635 \pm 0.02$. Our simulations even allow to see the variation of this ratio.

Finally, we plot on the left of Fig.~\ref{f:differences} the difference between the numerically obtained shape $\left< S(x)\right>$ and its  theoretically predicted value. On the right, we make the same comparison for the ratio $\left< S^2(x)\right>/ \left< S(x)\right>^2$. The precision achieved is a solid confirmation of our theory.

\section{Conclusions}
In this article, we considered the spatial shape of   avalanches at depinning. We gave scaling arguments showing that  close to the boundary in $d=1$, the averaged shape grows as a power law with the roughness exponent $\zeta$.  We then obtained analytically the full shape functions $\left< S(x)\right>_\ell $ for the BFM, where each degree of freedom sees a force which behaves as a random walk. 

It would be interesting to extend these considerations into several directions: First of all, one could ask what the shape function would be in higher dimensions. The techniques developed here will not immediately carry over: The domain where the advance of the avalanche is non-zero should be compact, but may have a fractal boundary. So we could still calculate the shape inside a given domain, but it would be meaningless to prescribe the boundary as  in $d=1$, where there are only two boundary points. 

Second, one can ask how the shape changes for short-range correlated disorder, by including perturbative corrections. Work in this direction is in progress. 

Finally, it would be interesting to obtain the avalanche shape for long-range elasticity, which is relevant for fracture, contact-line wetting, and earthquakes. The complication here is that an avalanche may contains several connected components.

\acknowledgements
We are grateful to Mathieu Delorme for providing the python code which generated the avalanches used in the numerical verification.

\appendix

\section{A   solution of $\tilde u''(x) + \tilde u(x)^2 = -\lambda \delta(x)$, with $\lambda \to -\infty$}
\label{a:instanton-solution-recall}
Let us  give a solution for the instanton equation with a single source \cite{DelormeLeDoussalWiese2016}, i.e.
\be\label{xxx}
\tilde u''(x) + \tilde u(x)^2 =- \lambda \delta(x)
\ .
\ee
The ansatz 
\be
\tilde u_{x_0}(x):=-\frac{6}{(|x|+x_0)^2}
\ee
  satisfies \Eq{xxx} with 
\be \label{A3}
-\lambda = \frac{24}{x_0^3}\ .
\ee
\begin{figure*}
\Fig{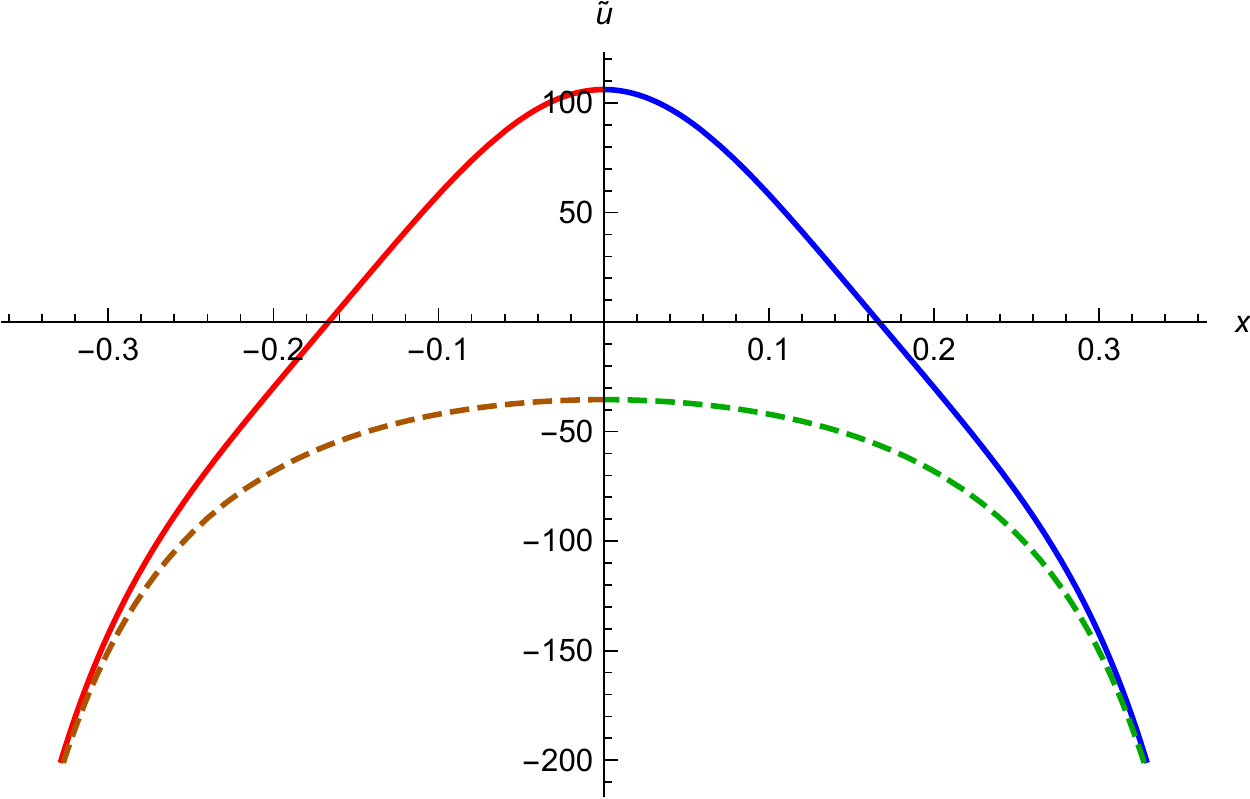}\Fig{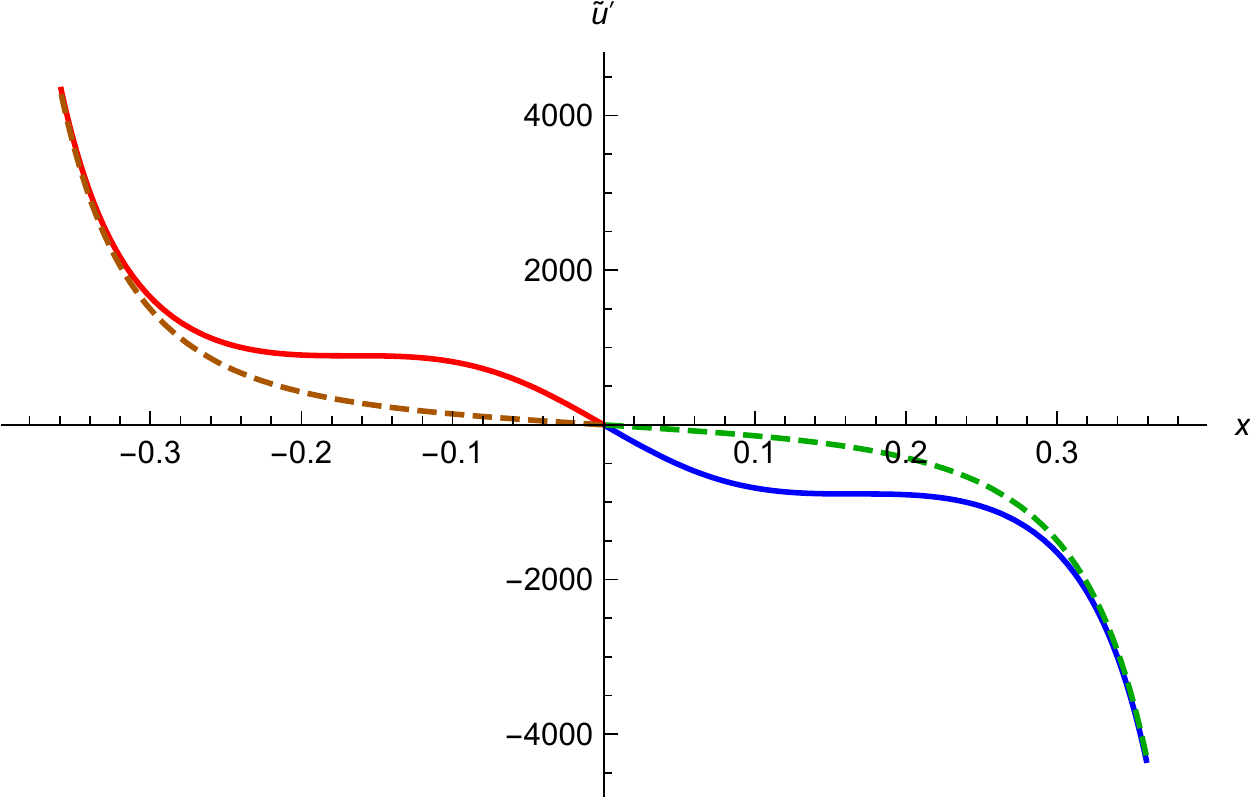}
\caption{The solutions $x(\tilde u)$, patching the two branches together at $x=0$, as well as its derivatives. In solid is the solution for ${\cal E}= {\cal E}_1>0$, in dashed the solution for $\bar{\cal E}= -\bar{\cal E}_1 <0$.}
\label{f:utilde0}
\end{figure*}
Note that this is an exact solution for a single source, but it also gives the leading behavior in case of several sources, especially how the non-trivial instanton-solution with two sources at $x=\pm 1/2$ can be regularized around its singularities. 

\section{Finite-energy instanton solutions}

We want to solve the instanton equation
\be
\tilde u''(x)+ \tilde u(x)^2 =0 \ .
\ee
Multiplying with $\tilde u'(x)$ and integrating once gives 
\bea
&& \frac { \tilde u'(x)^2}2 + \frac{\tilde u(x)^3}{3} = \cal E \ .
\eea
Solving for $\tilde u'(x)$ yields 
\bea
&& \tilde u'(x)=\pm \sqrt{2 {\cal E} -\textstyle \frac23 \tilde u(x)^3}\ , \\
&& \frac{\tilde u'(x)}{ \sqrt{2 {\cal E} -\textstyle \frac23 \tilde u(x)^3}} = \pm 1 \ .
\eea
Integrating once, we find
\bea
&& \frac{\tilde u \,
   _2F_1\left(\frac{1}{3},\frac{1}{2};\frac{4}
   {3};\frac{\tilde u^3}{3
   \mathcal{E}}\right)}{\sqrt{2 \mathcal{E}}} = \pm x + \mbox{const}\  .
\eea
These solutions are real for ${\cal E}>0$, which we consider first. 
\be
x_{\rm c}:=\lim_{u\to -\infty}
\frac{\tilde u \,
   _2F_1\left(\frac{1}{3},\frac{1}{2};\frac{4}
   {3};\frac{\tilde u^3}{3
   \mathcal{E}}\right)}{\sqrt{2 \mathcal{E}}} = -\frac{\sqrt[3]{3} \Gamma
   \left(\frac{1}{6}\right) \Gamma
   \left(\frac{4}{3}\right)}{\sqrt{2 \pi }
   \sqrt[6]{\mathcal{E}}}\ .
\ee
The solution stops at last argument of the hypergeometric function being 1, i.e.\ $u=\sqrt[3]{3 \cal E}$, s.t.
\be
x_0 :=  \frac{\tilde u \,
   _2F_1\left(\frac{1}{3},\frac{1}{2};\frac{4}
   {3};\frac{\tilde u^3}{3
   \mathcal{E}}\right)}{\sqrt{2 \mathcal{E}}}  \Big|_{u\to \sqrt[3]{3 \cal E}} = \frac{\sqrt[3]{3} \sqrt{\frac{\pi }{2}} \Gamma
   \left(\frac{4}{3}\right)}{\sqrt[6]{\mathcal{E
   }} \,\Gamma \left(\frac{5}{6}\right)}\ .
\ee
Note that 
$
x_{\rm c} = -2 x_0.
$
This allows us to write a solution symmetric around $x=0$,  (with the r.h.s. being positive)
\be
\pm x = \frac{\sqrt[3]{3} \sqrt{\pi } \Gamma
   \left(\frac{4}{3}\right)}{\sqrt{2}
   \sqrt[6]{ {\cal E}} \Gamma
   \left(\frac{5}{6}\right)}-\frac{\tilde u \,
   _2F_1\left(\frac{1}{3},\frac{1}{2};\frac{4}
   {3};\frac{\tilde u^3}{3
    {\cal E}}\right)}{\sqrt{2 {\cal E}}}\ .
\ee
The instanton has extension 1 for 
\be
{\cal E}_{1}:= \left( \frac{\sqrt[3]3\sqrt{2\pi}\Gamma(\frac 13)}{\Gamma(\frac 56)}\right)^{\!6} = \frac{52488 \pi ^3 \Gamma
   \left(\frac{4}{3}\right)^6}{\Gamma
   \left(\frac{5}{6}\right)^6}\ .
\ee
This  yields for the positive branch of the solution with extension  1 
\be
\pm x = \frac16 -\frac{{\tilde u} \,
   _2F_1\!\left(\frac{1}{3},\frac{1}{2};\frac{4}
   {3};\frac{{\tilde u}^3}{3
    {\cal E}_1}\right)}{\sqrt{2 {\cal E}_1}}
\ .
\ee
Now we consider solutions for $\bar {\cal E}:=-{\cal E} >0$. Using  Pfaffian transformations for the hypergeometric function yields
\be
\pm x = \frac{\sqrt{6} {\tilde u} \,
   _2F_1\left(\frac{1}{2},1;\frac{7}{6};\frac{
   3 \bar {\cal E}}{{\tilde u}^3+3
   \bar {\cal E}}\right)}{\sqrt{-3
   \bar {\cal E}-{\tilde u}^3}}+\frac{\sqrt[3]{3} \sqrt{2
   \pi } \Gamma
   \left(\frac{7}{6}\right)}{\sqrt[6]{\bar {\cal E}} \Gamma \left(\frac{2}{3}\right)}
\ .
\ee
Note that this solution is real; the shift  brings the solution around $x=0$. 
It has extension 1 in $x$-direction for 
\be
\bar{\cal E}_1=\left[ \frac{\sqrt{2 \pi } \Gamma
   \left(\frac{1}{3}\right)}{\sqrt[6]{3}
   \Gamma \left(\frac{5}{6}\right)}\right]^6 =\frac{8 \pi ^3 \Gamma
   \left(\frac{1}{3}\right)^6}{3 \Gamma
   \left(\frac{5}{6}\right)^6}\ .
\ee
There, 
\be
\pm x = \frac{\sqrt{6} {\tilde u} \,
   _2F_1\left(\frac{1}{2},1;\frac{7}{6};\frac{
   3 \bar {\cal E}}{{\tilde u}^3+3
   \bar {\cal E}}\right)}{\sqrt{-3
   \bar {\cal E}-{\tilde u}^3}}+\frac12 \Bigg|_{\bar {\cal E}=\bar {\cal E}_1}\ .
\ee
As is easily checked numerically, 
it agrees   with the  solution (52) of \cite{DelormeLeDoussalWiese2016}
\be
\tilde u(x) = - 6 \,{\cal P}\left(x+1/2; g_2=0,g_3 = \frac{\Gamma \left(\frac{1}{3}\right)^{18}}{
   (2\pi) ^6}\right) \label{50bis}\ .
\ee
\begin{figure*}
\Fig{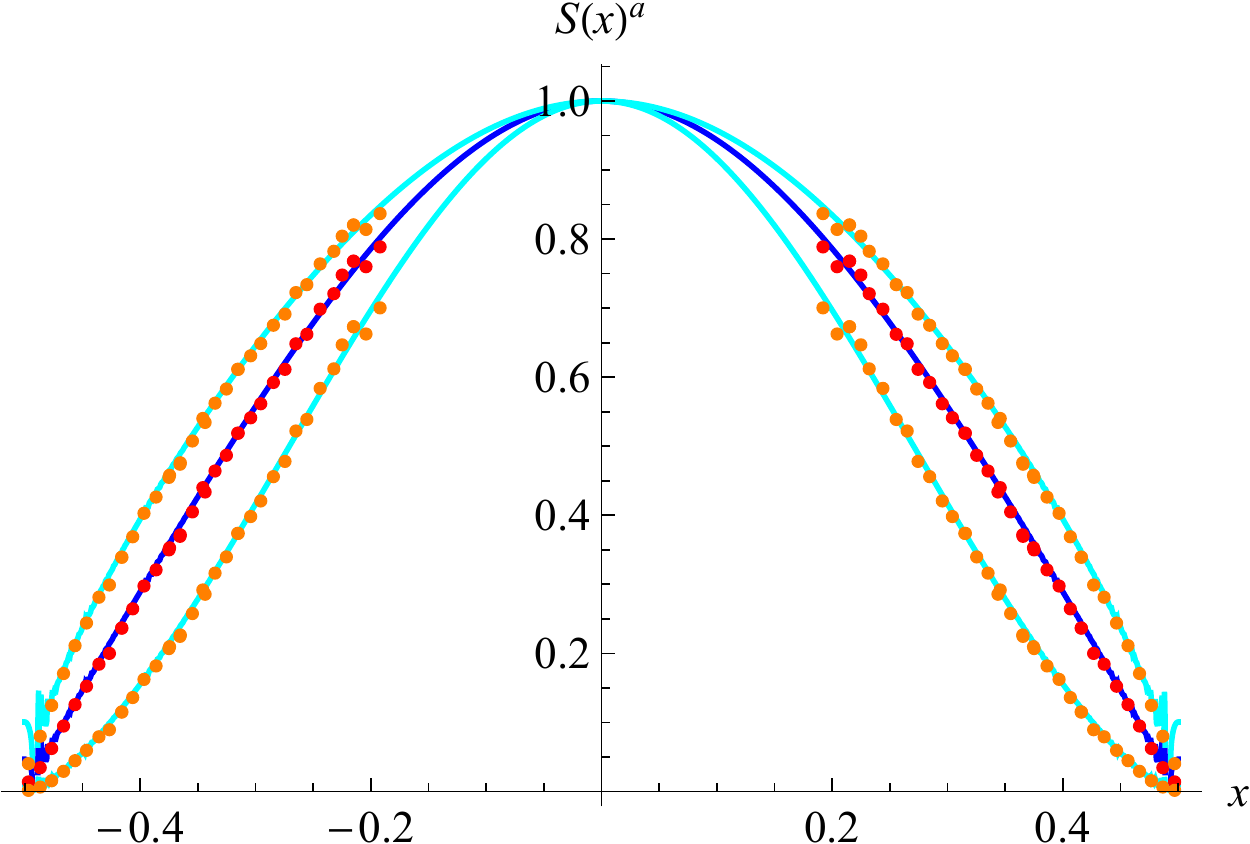}~~~~~~~\Fig{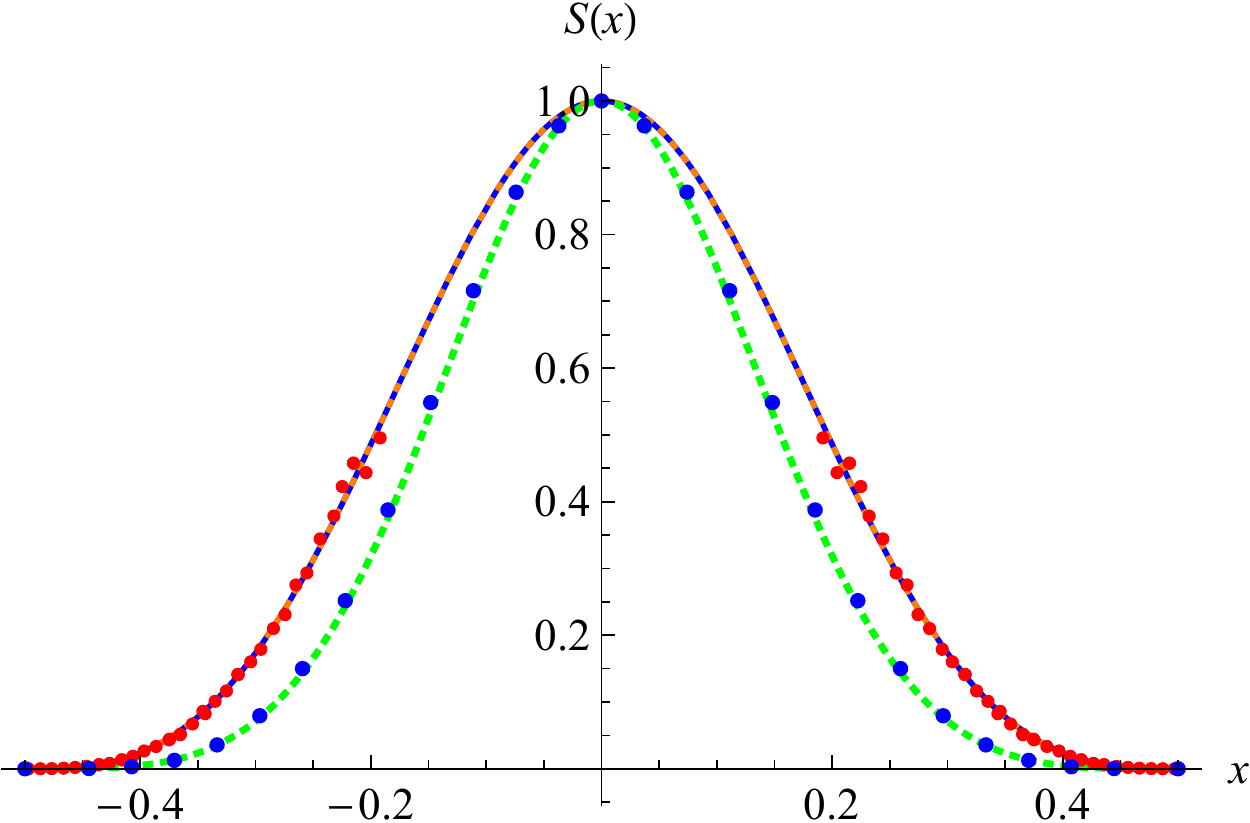}
\caption{Left: Data reverse engineered from \cite{ThieryLeDoussalWiese2015}, slightly shifted in $x$-direction and rescaled in $y$-direction to collapse with our result for $S(x)$, normalized to 1. The exponents from top to bottom are $a=1/4$, $a=1/3$, and $a=1/2$. Contrary to the claims of \cite{ThieryLeDoussalWiese2015}, $a=1/4$ is {\em not} the best fit, but $a=1/3$, corresponding to a cubic behavior at the boundary. Right: Consistency with our theory (top curve). This is compared to the theory in \cite{ThieryLeDoussalWiese2015} and its numerical validiation: The lower dashed curve is the theory for avalanches with a large aspect ratio $S/\ell^4$ while the dots are the numerical verification from the same reference.}
\label{f:data-Thimotee}
\end{figure*}The function ${\cal P}$ is the Weierstrass $\cal P$-function. By construction,  the solution $f(x) \equiv \tilde u(x) $ satisfies the following relations, which we give together for  convenience: 
\bea \label{44bis}
&&  {f^2(x) +f''(x) =0} \\
\label{45bis}
&&  {\frac23 f^3(x)+f'(x)^2 =-36  g_3 \equiv  -2 {\bar {\cal E}_1} } \\
\label{46bis}
&&  {\frac 23 f(x)f''(x)-f'(x)^2 = 36 g_3 = 2 \bar {\cal E}_1}
\eea
Using these relations, some terms which in general are not total derivatives can  be written as such, e.g.   
\be
f'(x)^2 = \frac{\rmd ^2}{\rmd x^2}\left[ \frac15 f(x)^2-\frac35 \overline{\cal E}_1 x^2 \right] 
\ .
\ee

\newpage

\section{Reanalysis of the data of Ref.~\cite{ThieryLeDoussalWiese2015}}
\label{a:replot}
In Ref.~\cite{ThieryLeDoussalWiese2015} it was claimed that   when averaging over all avalanches of a given extension $\ell$, close to the boundary the scaling function grows as $\left< S(x) \right> _\ell \sim (x-\ell/2)^4$. This was supported by a log-log plot of the data, see figure 14 of Ref.~\cite{ThieryLeDoussalWiese2015}. This procedure is dangerous, due to the boundary layer studied in section \ref{s:numerics}, which shifts the effective size of an avalanche. It is more robust to take $S(x)$ to the inverse expected power, and verify whether the resulting plot yields a straight line close to the boundary of the avalanche. This is done on Fig.\ \ref{f:data-Thimotee}. One clearly sees in the left plot that the data are most consistent with $a=\frac13$, equivalent to a cubic growth close to the boundaries. We also show in the right of Fig.\ \ref{f:data-Thimotee} that these data are consistent with our theory; note that the amplitude has been adjusted, since it could not be extracted from \cite{ThieryLeDoussalWiese2015}.

\tableofcontents

\vfill

\end{document}